\documentclass[floats,floatfix,amssymb,prd,twocolumn,superscriptaddress,nofootinbib]{revtex4-1}

\usepackage{amssymb,amsmath,verbatim,mathtools,needspace,enumitem,etoolbox,graphicx,physics,microtype,afterpage,bm}

\usepackage[dvipsnames, usenames]{xcolor}

\definecolor{linkcolor}{rgb}{0.0,0.3,0.5}
\usepackage[unicode, colorlinks=true, linkcolor=linkcolor, citecolor=linkcolor, filecolor=linkcolor,urlcolor=linkcolor, pdfusetitle]{hyperref}
\usepackage[all]{hypcap}
\usepackage[T1]{fontenc}
\usepackage[utf8]{inputenc}
\usepackage{tabularx}
\usepackage{float}
\usepackage{lmodern}

\allowdisplaybreaks

\usepackage{tikz}
\usepackage{color}
\usepackage{framed}
\usepackage{hyperref}
\hypersetup{colorlinks, citecolor=bluscuro, linkcolor=black, urlcolor=bluscuro}
\definecolor{rossos}{cmyk}{0,1,1,0.55}
\definecolor{bluscuro}{rgb}{0.15, 0.2, .85}
\definecolor{bluchiaro}{cmyk}{1,.3,0.,0.1}

\definecolor{ForestGreen}{rgb}{0.13, 0.55, 0.13}

\newcommand{\be}{\begin{equation}}
\newcommand{\ee}{\end{equation}}
\renewcommand{\d}{{\rm d}}
\def\BH{\text{\tiny BH}}
\newcommand{\llp}{\left [}
\newcommand{\rrp}{\right ]}
\newcommand{\lp}{\left (}
\newcommand{\rp}{\right )}

\def\lsim{\mathrel{\rlap{\lower4pt\hbox{\hskip0.5pt$\sim$}}
    \raise1pt\hbox{$<$}}}         
\def\gsim{\mathrel{\rlap{\lower4pt\hbox{\hskip0.5pt$\sim$}}
    \raise1pt\hbox{$>$}}}         

\begin{document}

\title{Tidal deformability of dressed black holes \\and tests of ultralight bosons in extended mass ranges}

\author{Valerio De Luca}
\affiliation{D\'epartement de Physique Th\'eorique and Centre for Astroparticle Physics (CAP), Universit\'e de Gen\`eve, 24 quai E. Ansermet, CH-1211 Geneva, Switzerland}
\affiliation{Dipartimento di Fisica, Sapienza Università 
di Roma, Piazzale Aldo Moro 5, 00185, Roma, Italy}

\author{Paolo Pani}
\affiliation{Dipartimento di Fisica, Sapienza Università 
	di Roma, Piazzale Aldo Moro 5, 00185, Roma, Italy}
\affiliation{INFN, Sezione di Roma, Piazzale Aldo Moro 2, 00185, Roma, Italy}

\date{\today}

\begin{abstract}
The deformability of a compact object under the presence of a tidal perturbation is encoded in the tidal Love numbers (TLNs), which vanish for isolated black holes in vacuum. 
We show that the TLNs of black holes surrounded by matter fields do not vanish and can be used to probe the environment around binary black holes. In particular, we compute the TLNs for the case of a black hole surrounded by a scalar condensate under the presence of scalar and vector tidal perturbations, finding a strong power-law behavior of the TLN in terms of the mass of the scalar field. Using this result as a proxy for gravitational tidal perturbations, we show that future gravitational-wave detectors like the Einstein Telescope and LISA can impose stringent constraints on the mass of ultralight bosons that condensate around black holes due to accretion or superradiance. Interestingly, LISA could measure the tidal deformability of dressed black holes across the range from stellar-mass ($\approx 10^2 M_\odot$) to supermassive ($\approx 10^7 M_\odot$) objects, providing a measurement of the mass of ultralight bosons in the range $(10^{-17} - 10^{-13}) \, {\rm eV}$ with less than $10\%$ accuracy, thus filling the gap between other superradiance-driven constraints coming from terrestrial and space interferometers. Altogether, LISA and Einstein Telescope can probe tidal effects from dressed black holes in the combined mass range $(10^{-17} - 10^{-11}) \, {\rm eV}$.
\end{abstract}

\maketitle

\section{Introduction}
Self-gravitating objects may get deformed when immersed in external tidal fields. Their deformability is measured in terms of the ``gravitational susceptibility'', also known as tidal Love numbers~(TLNs)~\cite{1999ssd..book.....M, poisson_will_2014}. The TLNs depend on the internal structure of the deformed body and leave detectable imprints in gravitational-wave~(GW) signals emitted during the last stages of a coalescence~\cite{Flanagan:2007ix}.

The relativistic TLNs of a neutron star have been computed in the seminal work by Hinderer~\cite{Hinderer:2007mb} and play a crucial role in GW astronomy, especially given the possibility to detect further neutron-star mergers and possibly mixed black hole-neutron star binaries in future observation runs by the LIGO-Virgo collaboration and by future instruments.
In particular, measuring the TLNs from a binary neutron-star coalescence provides stringent constraints on the equation of state~\cite{Baiotti:2010xh,Baiotti:2011am,Vines:2011ud,Pannarale:2011pk,Vines:2010ca,Lackey:2011vz,Lackey:2013axa,
  Favata:2013rwa,Yagi:2013baa,Maselli:2013mva,Maselli:2013rza,DelPozzo:2013ala,TheLIGOScientific:2017qsa,Bauswein:2017vtn, Most:2018hfd,Harry:2018hke,Annala:2017llu, Abbott:2018exr,Akcay:2018yyh,Abdelsalhin:2018reg, Jimenez-Forteza:2018buh, Banihashemi:2018xfb, Dietrich:2019kaq, Dietrich:2020eud, Henry:2020ski, Pacilio:2021jmq} (see
Refs.~\cite{GuerraChaves:2019foa,Chatziioannou:2020pqz} for some recent reviews).

In parallel, the TLNs have recently acquired considerable attention within the gravity-theory and the high-energy physics communities, due to the peculiar tidal deformability of black holes~(BHs). An intriguing result in general relativity is that the TLNs of \emph{naked} BHs (i.e. those in isolation) is found to be exactly zero, see Refs.~\cite{
Binnington:2009bb,Damour:2009vw,Damour:2009va,Pani:2015hfa,Pani:2015nua,Gurlebeck:2015xpa,Porto:2016zng,
LeTiec:2020spy, Chia:2020yla,LeTiec:2020bos,Hui:2020xxx,Charalambous:2021kcz,Charalambous:2021mea} for literature on this topic. 
The vanishing of the BH TLNs poses a naturalness problem from the effective-field-theory point of view~\cite{Porto:2016zng} and is in fact a fragile property, which is broken for any other compact object~\cite{Cardoso:2017cfl}, in higher dimensions~\cite{Kol:2011vg,Cardoso:2019vof, Hui:2020xxx}, in modified gravity~\cite{Cardoso:2017cfl,Cardoso:2018ptl}, and is indeed related to special symmetries emerging in the four-dimensional, vacuum, general-relativistic case~\cite{Hui:2020xxx,Charalambous:2021kcz,Charalambous:2021mea,Hui:2021vcv}.

A natural question which arises at this point is if this property is maintained also for \emph{dressed} BHs, i.e. those surrounded by matter fields. The latter can grow around BHs due to secular effects like accretion or due to more dramatic processes like superradiant instabilities~\cite{Zeldovich:1982zz,Press:1972zz, Teukolsky:1974yv,Arvanitaki:2014wva,Arvanitaki:2016qwi} (see Ref.~\cite{Brito:2015oca} for a review).
It is reasonable to expect that dressed BHs would have nonvanishing TLNs. Depending on their magnitude, the latter can impact the GW signal of a binary BH coalescence in the latest stage before the merger and provide a smoking gun for some ``structure'' near merging BHs. In fact, any observation of a sufficiently massive coalescing binary with a nonvanishing tidal deformability would signal some departure from the standard ``naked BH'' paradigm. 

The scope of this work is to quantify this effect. Accreting BH solutions can be described in terms of generalisations of the Vaidya spacetime~\cite{Vaidya:1951zz, Vaidya:1953zza, Vaidya:1999zza, Stachel:1977cm, Bonnor:1970zz}. In addition, in the presence of ultralight bosonic degrees of freedom~\cite{Hui:2016ltb}, macroscopic quasi-stationary condensates can grow around spinning BHs on time scales much shorter than the typical accretion time scale. While the self-gravity effects of ordinary/dark matter halos and accretion structures have a small impact during the inspiral~\cite{Barausse:2014tra}, if sufficiently dense (as in the case of boson condensates discussed below) they can survive during the late inspiral before being tidally disrupted, affecting the GW signal through tidal-deformability effects.

We shall mostly focus on the case of massive scalar fields which can grow around BHs either through accretion or due to the BH superradiant instability if their Compton wavelength is comparable to the BH radius (which requires the field to be ultralight). In the latter case, the low-energy bosons can efficiently extract the rotational energy of a BH over relatively short time scales~\cite{Brito:2015oca}. This happens as long as the superrandiance condition for the mode's frequency  $w < m \Omega$ is maintained, where $m$ is the azimuthal quantum number of the mode and $\Omega$ the angular velocity of the BH. In particular, the unstable mode will grow exponentially near the BH, extracting energy and angular momentum and forming a tidally-locked, rotating condensate around the BH until the superradiance condition is saturated; at this point the cloud will stop growing and the condensate will be slowly dissipated in GWs.

Striking features of the BH superradiant instability triggered by ultralight bosons include gaps in the spin-mass distribution of 
astrophysical BHs~\cite{Arvanitaki:2014wva,Arvanitaki:2016qwi,Stott:2020gjj}, a continuous GW signal emitted by the condensate~\cite{Arvanitaki:2014wva,Brito:2017wnc, Brito:2017zvb}, a stochastic GW background generated by unresolved sources~\cite{Brito:2017wnc,Brito:2017zvb}, and 
self-gravity~\cite{Hannuksela:2018izj} and tidal effects~\cite{Baumann:2018vus,Zhang:2018kib,Berti:2019wnn,Baumann:2019eav,Baumann:2019ztm,Zhang:2019eid,Cardoso:2020hca} in binary inspirals. Negative searches for continuous GW signals~\cite{Palomba:2019vxe, Zhu:2020tht} and for the stochastic background generated by unresolved sources~\cite{Brito:2017wnc, Brito:2017zvb, Tsukada:2018mbp, Tsukada:2020lgt} with the LIGO/Virgo interferometers have set constraints on the mass of the ultralight bosons in a narrow range around $\sim 10^{-13}\, {\rm eV}$.
Absence of mass-spin gaps places wider, although indirect, constraints~\cite{Stott:2020gjj}. Future experiments like LISA can potentially investigate masses in the range $(10^{-19} - 10^{-15}) \, {\rm eV}$~\cite{Brito:2017wnc, Brito:2017zvb,Brito:2020lup}. Multiband searches with ground- and space-based detectors~\cite{Ng:2020jqd} and measurements of the stochastic background with third-generation instruments~\cite{Yuan:2021ebu} will slightly enlarge these mass ranges. Altogether, electromagnetic and GW observations can potentially constrain the mass $m_b$ of a putative ultralight bosonic field in a wide range $10^{-22} \lesssim m_b c^2\,{\rm /eV} \lesssim 10^{-10}$, although an interesting one-order of magnitude wide window, approximately around $10^{-14}\,{\rm eV}$, remains hard to probe (see Ref.~\cite{Brito:2015oca} for a recent summary of the constraints).

As we shall show, measurements of the tidal deformability of BH-boson condensates with future detectors can potentially probe a mass range that is complementary to other tests, thus filling the gap between other superradiance-driven constraints coming from terrestrial and space interferometers.

The rest of this paper is organized as follows. In Sec.~II we present our framework to describe a BH surrounded by a massive scalar field. This configuration will then be considered as the background to study, in Sec.~III, the Love numbers for scalar and vector tidal perturbations. In Sec. IV we investigate the detectability of this effect whereas in Sec.~V we summarize our findings. Throughout this paper we use geometrical units ($G = c = 1$). Appendix~A is devoted to the computation of the Love numbers for a neutron star under the presence of scalar and vector tidal perturbations.

\section{Framework}

Let us start by considering a spherically symmetric isolated body possibly absorbing matter from a reservoir placed at large distances. The system is assumed to be in equilibrium, where the loss of matter into the BH (if any) is balanced by the infall of matter from infinity, forming a quasi-stationary flow.  The Einstein equations, with the matter fields determining the stress-energy tensor $T_{ab}$, are given by $G_{ab}= 8 \pi T_{ab}$, where $G_{ab}$ is the Einstein tensor of the metric.

Following the perturbative scheme developed in Refs.~\cite{Babichev:2012sg, Bamber:2021knr}, one can solve the problem perturbatively assuming the matter fields are a small perturbation of a naked BH. The latter is simply described by the Schwarzschild metric, which 
can be expressed in terms of the ingoing Eddington-Finkelstein coordinates
$ v = t + r_*= r + r_\BH {\rm ln} (r/r_\BH - 1)$ as
\be
\d s^2 =  - f \d v^2 + 2 \d v \d r + r^2 \d \Omega^2, \qquad f (r) = 1 - \frac{2  M_\BH}{r},
\ee
where we have defined the BH Schwarzschild radius $r_\BH = 2 M_\BH$.

\subsection{Massive scalar perturbations}

We will now focus on the case in which the matter source is determined by a (real or complex\footnote{As discussed below, in our context the notable difference between real and complex bosonic condensates
is that the latter, in a nearly stationary regime, have a constant-in-time stress–energy tensor, and therefore do not emit GWs.}) scalar field $\Phi$ with mass $m_b=\mu \hbar$, 
satisfying the Klein-Gordon equation $\Box \Phi = \mu^2 \Phi$. To linear order in the matter-field perturbations, the Klein-Gordon equation should be evaluated on the Schwarzschild metric.

The background scalar field can be decomposed in terms of the spin-0 spheroidal\footnote{Spheroidal harmonics reduce to the standard spherical harmonics in the case of nonspinning BHs. Since later on we will discuss the superradiant instability of spinning BHs, for the sake of generality we use the spheroidal harmonics. As we shall discuss, in the small-$\alpha$ limit the spin of the BH can be neglected when computing the eigenfunctions~\cite{Brito:2014wla}.} harmonics $S_{\ell m} (\theta)$ as~\cite{Brito:2014wla}
\be
\Phi = \int \d w \, \sum_{\ell m} e^{-i w v} e^{i m \varphi} S_{\ell m} (\theta) \psi_{\ell m} (r).
\ee
In the limit of small gravitational coupling, $\alpha \equiv M_\text{\tiny BH}\mu$, given by the ratio of the BH horizon radius and the reduced Compton wavelength of the scalar field,  the radial part of the wavefunction can be approximated by~\cite{Yoshino:2013ofa,Brito:2014wla}
\be
\psi (r) = \Phi_0  \lp \frac{4\alpha^2}{\ell+1} \frac{r}{r_\BH} \rp^\ell \exp\lp -\frac{2\alpha^2}{\ell+1} \frac{r}{r_\BH}\rp,
\ee
where $\Phi_0$ is an arbitrary amplitude, we have focused on the fundamental ($n=0$ node) mode, and defined $\psi\equiv \psi_{\ell m}$ for simplicity. 
The corresponding eigenfrequency is complex, $w=w_R+iw_I$, with $w_R\sim \mu$ and $w_I \sim- w_R \alpha^{4\ell+5}$~\cite{Brito:2015oca}.  
Note that the radial eigenfunction peaks at $r_s = r_\BH(\ell+1)^2/2\alpha^2$~\cite{Arvanitaki:2010sy}, which is much larger than the BH horizon for small values of $\alpha$. Indeed, the eigenfunctions can be also computed in the flat spacetime approximation~\cite{Yoshino:2013ofa,Brito:2014wla}.

Note that a complex scalar field circumvents the hypotheses of the no-hair theorems~\cite{Chrusciel:2012jk, Herdeiro:2015waa}, giving rise to stationary hairy BHs. On the other hand, in the $\alpha\ll1$ limit, a real scalar field supports long-lived, quasi-bound states that slowly dissipate through GW emission. The time scale for this process is very long so for our purposes we can assume the corresponding hairy BH is nearly stationary.

\subsection{Metric backreaction}
\subsubsection{Spherical case}
Let us first consider the spherically symmetric ($\ell=0$) case. In general, the perturbation to the metric due to the effect of a small spherically symmetric accretion flow can be expressed as~\cite{Babichev:2012sg, Bamber:2021knr}
\be
\d s^2 =  - F e^{2 \delta \lambda (v, r)} \d v^2 + 2 e^{\delta \lambda (v, r)} \d v \d r + r^2 \d \Omega^2,
\ee
in terms of the perturbation of the metric $\delta \lambda (v, r)$ and 
\be
F (v,r) = f(r) - \frac{2 \delta M (v, r)}{r} = 1 - \frac{2  M_\BH + 2 \delta M (v, r)}{r}.
\ee
The mass perturbation $\delta M (v, r)$ describes the additional contribution to the BH mass due to the presence of accretion.

Assuming that accretion onto the BH occurs from an asymptotically constant energy density $\rho = - T^v_v$,
one can solve the perturbed Einstein equations to find~\cite{Babichev:2012sg, Bamber:2021knr}
\begin{align}
\delta M (v,r) & = \delta \tilde M \, v + \int_{r_\BH}^r 4 \pi r'^2 \rho \d r', \nonumber \\
\delta \lambda (v,r) & = - \int_{r_\BH}^r 4 \pi r' T_{rr} \d r',
\end{align}
in terms of the rate of increase in the mass of the BH, $\delta \tilde M$, which depends on the particular field configuration (see below).
We stress that this perturbative scheme is valid as long as these corrections are small, which means that
the total flux of the infalling matter should be sufficiently small such that the energy density of the matter is small compared to $\sim1/M_\BH^2$ and the mass of matter inside a radius $r$ is much smaller than the bare BH mass.

In the particular case of a spherically symmetric (real or complex) massive scalar field described above, the energy density is given by~\cite{Bamber:2021knr}
\be
\rho = f |\partial_r \Phi|^2 + \mu^2 |\Phi|^2,
\ee
and has a regular behavior at the horizon as $\rho \to \mu^2 |\Phi_0|^2$.
The backreaction to the metric takes the explicit form~\cite{Bamber:2021knr}
\begin{align}
\delta M (v,r) & = 8 \pi (2 M_\BH w)^2 |\Phi_0|^2 v \nonumber \\
& + \int_{r_\BH}^r 4 \pi r'^2 (f |\partial_r \Phi|^2 + \mu^2 |\Phi|^2) \d r', \nonumber \\
\delta \lambda (r) & = - 2\int_{r_\BH}^r 4 \pi r' |\partial_{r} \Phi|^2 \d r'. 
\end{align}
One can notice that, at leading order in the matter perturbations, the horizon of the accreting BH increases as
\be
r_\BH^\text{\tiny acc} (v) = r_\BH + 64 \pi |\Phi_0|^2 M_\BH^2 \mu^2 v, \label{rH}
\ee
being a function of the advanced-time coordinate $v$.

\begin{figure*}[t!]
	\centering
	\includegraphics[width=0.48 \linewidth]{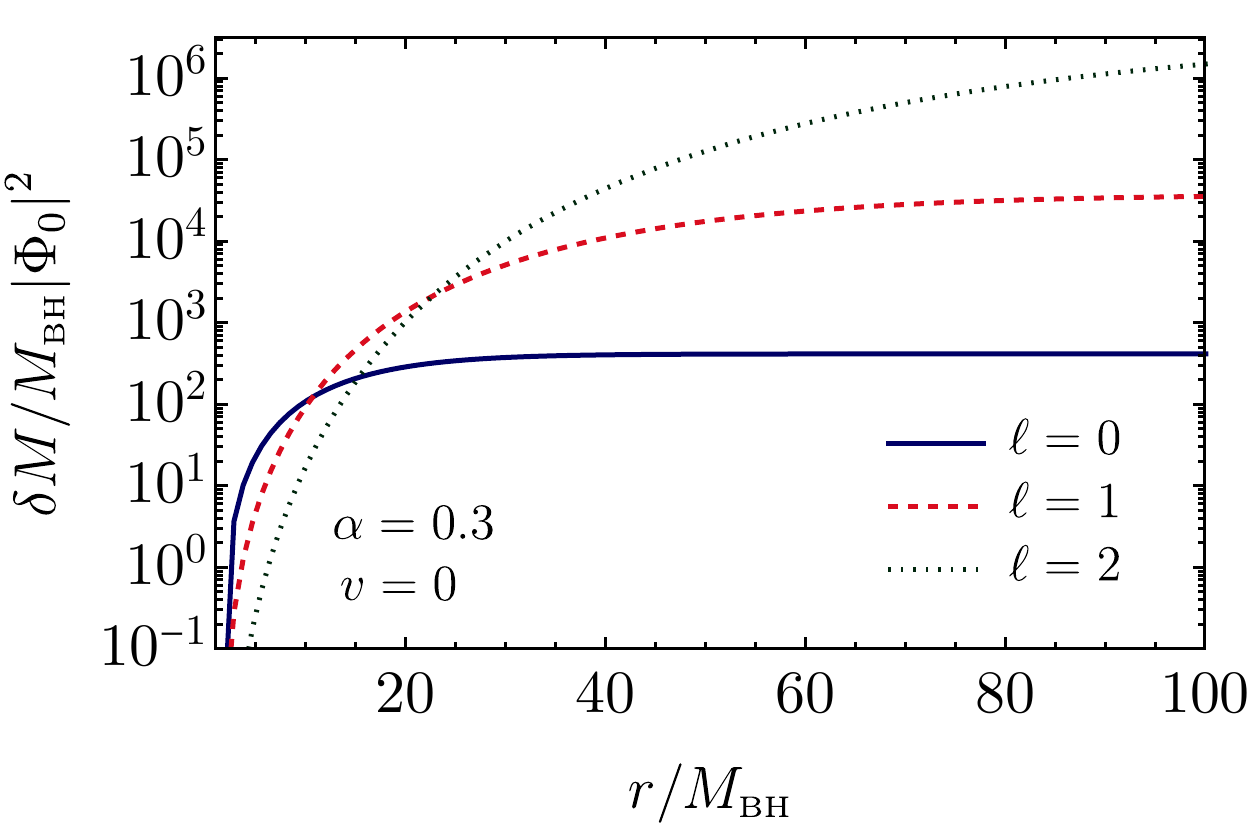}
	\includegraphics[width=0.48 \linewidth]{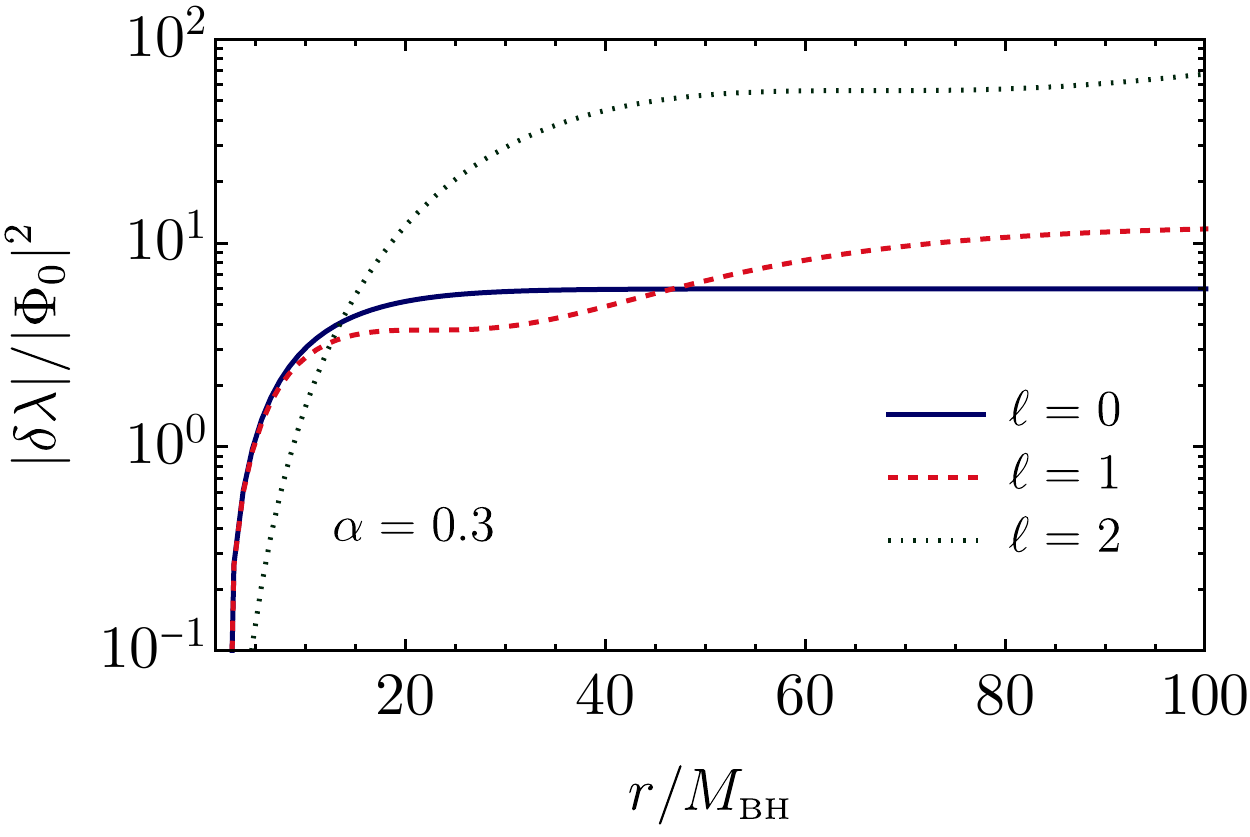}
	\caption{\it Profile of the corrections to the mass $\delta M$ and to the metric $\delta \lambda$ in terms of the radial coordinate, for a fixed value of the gravitational coupling $\alpha = 0.3$ and for the modes $\ell=0$, $\ell=1$  and $\ell=2$ of the complex scalar field background.
	}
	\label{fig:1}
\end{figure*}

In Fig.~\ref{fig:1} we show the radial dependence of the mass $\delta M$ and metric $\delta \lambda$ perturbation for fixed value of the gravitational coupling and for the $\ell=0$ mode, along with the $\ell=1,2$ modes discussed below. One can easily appreciate that both corrections go to zero in the limit $r \to r_\BH$ (assuming $v=0$ for the mass correction), and that higher multipoles of the scalar field will result in larger perturbations of the background metric (but assuming the same amplitude for the $\ell=1,2$ modes, which is typically not the case as discussed below). Finally, given the exponential suppression of the massive scalar field at large radii, both corrections approach a plateau for $r \gg r_s$.

The amplitude of the scalar field can be expressed in terms of the mass of the scalar cloud $M_s$ as~\cite{Brito:2014wla}
\be
M_s  = 4\pi \int_{r_\BH}^\infty \d r \, r^2  \rho = 4\pi \int_{r_\BH}^\infty \d r \, r^2 \lp f |\partial_r \Phi|^2 + \mu^2 |\Phi|^2 \rp,
\ee
such that 
\begin{align}
 |\Phi_0^{\ell=0}|^2 &= \frac{1}{\pi} \lp \frac{M_s}{M_\BH} \rp \alpha^4 e^{4 \alpha^2}. \label{Phi0}
\end{align}
The energy density in this case reads
\begin{align}
 \rho &=\frac{\alpha^6}{\pi M^2_\BH} \frac{M_s}{M_\BH} \nonumber \\
 &\sim 2\times 10^{-2} \left(\frac{\alpha}{0.1}\right)^6\left(\frac{10^6 M_\odot}{M_\BH}\right)^2 \frac{M_s}{0.1 M_\BH}\, {\rm g}/{\rm cm}^3\,.\label{rhonorm}
\end{align}
When $\alpha\ll1$, $\rho$ is much smaller than the typical BH energy density, $\rho_\BH M^2_\BH=3/(32\pi)\sim {\cal O}(0.1)$, even when $M_s$ is a sizeable fraction of the BH mass. As a reference, the typical density of a thin accretion disk around a $10^6 M_\odot$ BH is $\rho_{\rm disk}\approx 6\times 10^{-3}f_{\rm Edd}^{11/20}\, {\rm g}/{\rm cm}^3$~\cite{1973A&A....24..337S} (where $f_{\rm Edd}$ is the Eddington fraction for mass accretion), i.e. it is comparable to Eq.~\eqref{rhonorm} with the chosen normalization. Based on this observation, later on we shall use $M_s\sim 0.1 M_\BH$ as our fiducial value, which also saturates the condition $M_s\lesssim 0.1M_\BH$ beyond which the perturbative scheme of Refs.~\cite{Babichev:2012sg, Bamber:2021knr} breaks down.

For the relevant case of a  binary system during its inspiral phase, one can note that the characteristic accretion time scale, $\tau_\text{\tiny Salpeter}\simeq 4.55\times 10^7 \,{\rm yr}$,
is much longer than the orbital period,
\be
T = \sqrt{\frac{4 \pi^2 a^3}{2 M_\BH}} \sim 6 \times 10^{-2} \, {\rm s} \lp \frac{\beta}{0.1}\rp^{-3} \lp \frac{M_\BH}{M_\odot} \rp,
\ee
where $a$ is the semi-major axis of the binary and $\beta$ is the orbital velocity. Therefore, during the inspiral, one can assume that accretion is an adiabatic process and neglect the time dependence of the metric perturbation. In practice, one can assume that $\delta M (v,r) \sim \delta M(\tilde v, r)$, with $\tilde v$ approximately constant during the inspiral time scale. 
As we shall show, in this regime the parameter $\tilde v$ does not affect the TLNs.

\subsubsection{Superradiantly-triggered bosonic condensate}

While spherical condensates can only grow through accretion, nonaxisymmetric modes of ultralight fields can trigger the formation of a macroscopic bosonic condensate around spinning BHs by the superradiant instability~\cite{Brito:2015oca}.

The BH  superradiant instability is satisfied by modes with
$w_R < m \Omega$, so it requires nonaxisymmetric modes with azimuthal number $m$ and a spinning BH with angular velocity $\Omega = \chi/(2 M_\BH (1+\sqrt{1-\chi^2}))   $, where $\chi$ is the dimensionless BH spin.

The exponential growth of the $\ell=m=1$ unstable mode around a BH from the extraction of its energy and angular momentum proceeds within a characteristic time scale~\cite{Detweiler:1980uk}
\be
	\tau_I \sim \frac{1}{w_I}\sim 7 \times  10^{-3} \chi^{-1} \lp \frac{M_\BH}{M_\odot} \rp \lp \frac{\alpha}{0.1} \rp^{-9} {\rm yr}.
\ee
 As one can appreciate, this time scale is shorter than the one relevant  for accretion, $\tau_\text{\tiny Salpeter}$. See in particular Ref.~\cite{Brito:2014wla} for a detailed analysis on the interplay between these phenomena.
This instability will proceed as long as the condition for the superradiant instability is satisfied
\be
\chi >\chi_\text{\tiny crit} \simeq \frac{4 m \alpha}{m^2 + 4 \alpha^2}\,.
\ee
When this condition is saturated, the cloud will stop growing and will start dissipating by emitting GWs\footnote{This occurs only for real bosonic fields. A complex massive bosonic field satisfying the condition $w=m\Omega<\mu$ does not dissipate at the horizon (since the horizon flux is proportional to the superradiant factor $w-m\Omega$, and vanishes at the ``tidal locking'' condition $w=m\Omega$), nor at infinity (since modes with $w<\mu$ are Yukawa-suppressed and do not propagate) and can form stationary BHs known as Kerr BHs with bosonic hair~\cite{Herdeiro:2014goa,Herdeiro:2015waa}. For our purposes these solutions can be discussed as in the case of real fields, but neglecting their GW emission~\cite{Brito:2014wla}. } in a time scale of the order of
\be
	\tau_\text{\tiny GW} \sim 6 \times  10^{3} \chi^{-1} \lp \frac{M_\BH}{M_\odot} \rp \lp \frac{\alpha}{0.1} \rp^{-15} {\rm yr},
\ee
which is much longer than the instability time scale, $\tau_I$. Therefore, in the $\alpha\ll1$ regime, the process can be conveniently studied in two steps: first the condensate grows on a time scale $\tau_I$ until saturation and, then, it dissipates in GWs over a time scale $\tau_\text{\tiny GW}\gg \tau_I$.

The extraction of the spin of a BH in a superradiance process is accompanied by 
a decrease of the BH mass according to the relation
\be
\Delta M_\BH\sim M_s \sim  -\mu \llp \chi_f M_\BH^2 - \chi_i (M_\BH^{i})^2 \rrp,
\ee
in terms of the initial (final) BH mass and spin $M_\BH^{i}$, $\chi_i$ ($M_\BH$, $\chi_f$). Solving for $M_s$, one finds
\be
M_s = \frac{1-2 M_\BH \chi_i \mu - \sqrt{1- 4  \chi_i M_\BH \mu + 4  \chi_i  \chi_f M_\BH^2 \mu^2}}{2 \chi_i \mu},
\ee
and using the superradiance condition
\be
\chi_f = \chi_\text{\tiny crit} \simeq \frac{4}{m} \alpha + \mathcal{O}(\alpha^3), \label{chif}
\ee
at the leading order in the gravitational coupling $\alpha$, one finds
\be
\frac{M_s}{M_\BH} = \frac{1-2 \chi_i \alpha - \sqrt{1- 4 \chi_i \alpha + 16 \chi_i \alpha^3}}{2 \chi_i \alpha} \simeq \chi_i \alpha + \mathcal{O}(\alpha^2).
\ee
For an initially highly spinning BH, one finds
\be
\frac{M_s}{M_\BH}  \simeq \alpha,
\ee
which provides an estimate for the final mass of the scalar condensate around the BH. 

In the numerical analysis presented below we shall assume the upper bound on the scalar cloud mass $M_s = 0.1 M_\BH$, which roughly corresponds to the maximum superradiant energy extraction~\cite{Brito:2015oca,Begelman:2014aea,Ficarra:2018rfu}, and also saturates the validity regime of the perturbative expansion.

Note that, at the end of the superradiant process, the BH is (slowly) spinning and the condensate is non-axisymmetric. For example, for the dominant $\ell=m=1$ mode of a real scalar field one has~\cite{Brito:2014wla}
\begin{equation}
 \Phi \propto \psi(r) \cos(\varphi-w_R t)  \sin\theta\,.
\end{equation}
In the $\alpha\ll1$ limit, since the condensate peaks at $r_s\gg r_\BH$, the BH spin can be neglected and the scalar field can be decomposed in spherical (rather than spheroidal) harmonics, $\Phi\sim \psi(r)\sum_{\ell m} a_{\ell m} Y^{\ell m}$. For example, for the $\ell=m=1$ case above the only nonvanishing harmonics are $Y^{1\,\pm1}$. 
When plugged back into Einstein's equations, owing to the fact that $T_{\mu\nu}$ is quadratic in the scalar field and to the standard angular-momentum sum rules~\cite{Pani:2013pma}, an $\ell=1$ scalar mode will source $\ell=0,2$ modes in the metric perturbations of a naked Schwarzschild BH. The $\ell=0$ components can be treated as discussed above for spherical perturbations, whereas the $\ell=2$ modes will produce a quadrupole moment, $Q\propto M_s^2$, on the background.

At this stage, when dealing with superradiantly-triggered condensates, we make two approximations: i) we neglect the spin of the final BH, since it is anyway small [see Eq.~\eqref{chif}] and plays a minor role in the $\alpha\ll1$ regime; ii) we neglect the nonspherical backreaction of the condensate onto the metric. The first assumption is well justified for our purposes. Indeed, in the next section we will be interested in computing the TLNs of this background configuration. Schematically, the TLN of a slowly-spinning object can be written as~\cite{Pani:2015hfa,Pani:2015nua}
\begin{equation}
 {\rm TLN} = {\rm TLN}^{\rm nonspinning}\left(1 + a_1 m\chi_f   +{\cal O}(\chi_f)\right),
\end{equation}
where $a_1$ is an order-unity dimensionless coefficient. Since $m\chi_f\sim \alpha\ll1$, the leading-order contribution to the TLNs comes from the spherically-symmetric part of the metric, exactly like in the neutron-star case~\cite{Pani:2015nua}.
The second approximation above is instead a working assumption, since in principle the $\ell=1$ (say) condensate will source $\ell=0$ and $\ell=2$ metric perturbations with comparable amplitudes. However, for the purposes of estimating the order of magnitude of tidal effects, one can focus on the $\ell=0$ contribution only, which is much simpler to compute. The $\ell=2$ background perturbations will yield a TLN of the same order of magnitude, which should sum with the $\ell=0$ contribution. 
Neglecting the nonspherical part is also in line with the fact that in the next section we shall study only scalar and vector tidal fields and use the result as a proxy for the more relevant gravitational TLNs. Therefore, for what concerns the case of superradiantly-triggered nonaxisymmetric condensates, our computation should be considered as an order-of-magnitude estimate.

Regardless of the nonspherical contributions, one can compute the spherically-symmetric part of the metric as in the previous section. The result for the $\ell=1,2$ modes is shown in Fig.~\ref{fig:1} together with the (nonsuperradiant) $\ell=0$ mode previously discussed.
Neglecting the nonspherical metric backreaction, one can map the amplitude of the scalar field into the condensate mass as
\begin{align}
|\Phi_0^{\ell=1}|^2 &= \frac{1}{96 \pi} \lp \frac{M_s}{M_\BH} \rp \alpha^4 e^{2 \alpha^2},  \\
|\Phi_0^{\ell=2}|^2 &= \frac{1}{9720 \pi}  \lp \frac{M_s}{M_\BH} \rp \alpha^4 e^{\frac{4}{3} \alpha^2}.
\end{align}
In this last step we have assumed the simplified scenario where the mass of the scalar cloud is fully
accounted by just one dominant mode.
In particular, since the superradiant condition is roughly $\mu<m\Omega$, if it is satisfied for the mode $m=1$, it is also so for any $m>1$. However, the instability time scale for higher-$m$ modes is much longer, so by the time the dominant ($m=1$) mode has extracted sufficient angular momentum from the BH, the other modes did not have enough time to grow sufficiently~\cite{Ficarra:2018rfu}. The only case in which higher-$m$ modes are relevant is if the superradiant condition is \emph{not} satisfied for $m=1$ but only for higher values of $m$, which happens after the dominant mode has already saturated. In this case the next dominant mode becomes the superradiantly unstable one with the smallest value of $m$, producing a cascade process of ever longer time scales. See Ref.~\cite{Ficarra:2018rfu} for a detailed discussion on multiple modes.

\section{Scalar and Vector Love Numbers}
In this section we compute the TLNs for a BH surrounded by a massive scalar field condensate under the presence of an external scalar ($\phi$) and vector ($A_\mu$) tidal field. In order to do so, we report the main steps to perform this calculation. 

\subsection{Definition of scalar and vector Love numbers}
Focusing on a nonspinning object of mass $M$, one can extract the multipole moments from the asymptotic expansion of the relevant fields. For a gravitational tidal field one can extract the (electric) multipole moments from~\cite{Thorne:1980ru,Cardoso:2017cfl}
\begin{align}
\label{eq:gttexpansion}
g_{tt}&=1+\frac{2M}{r}+ \sum_{l\geq 2}\left(\frac{2}{r^{l+1}}\left[\sqrt{\frac{4\pi}{2l+1}}M_l Y^{l0}  + (l'<l\,\text{pole})\right] \right. \nonumber \\
& \left. - \frac{2}{l(l-1)}r^l\left[\mathcal{E}_l Y^{l0}+(l'<l\,\text{pole})\right]\right)\,,
\end{align}
where $M_l$ are the mass multipole moments ($l=2,3,..$, not to be confused with the $\ell$ index describing the multipolar structure of the background scalar condensate) and $\mathcal{E}_l$ is the amplitude of the multipole moment $l$ of the tidal field.

We can perform a similar asymptotic expansion for the (axial, or parity-odd) vector field $A_\varphi$ and for the scalar field $\phi$ as~\cite{Cardoso:2017cfl}
\begin{align}
\label{eq:Aphiexpansion}
A_\varphi &=\sum_ {l\geq 1} \left(\frac{2}{r^l}\left[\sqrt{\frac{4\pi}{2l+1}}J_l S_\varphi^{l0} + \left(l'<l\,\text{pole}\right)\right] \right. \nonumber \\
& \left. -\frac{2r^{l+1}}{l\left(l-1\right)}\left[\mathfrak{B}_l  S_\varphi^{l0}+\left(l'< l\,\text{pole}\right)\right]\right),\\
\label{eq:scalarexpansion}
\phi&=\phi_0 +\sum_{l\geq 1} \left(\frac{1}{r^{l+1}}\left[\sqrt{\frac{4\pi}{2l+1}}\phi_l Y^{l0} + (l'<l \,\text{pole})\right] \right. \nonumber \\
& \left. - \frac{1}{l(l-1)} r^l \left[\mathcal{E}^\text{S}_l+\left(l'< l\,\text{pole}\right)\right]\right),
\end{align}
in terms of the magnetic $J_l$ and scalar $\phi_l$ multipole moments, where we have defined $\mathfrak{B}_l$ as the axial (parity odd) vector tidal field, and $\mathcal{E}^\text{S}_l$ as the scalar one.

The TLNs due to a gravitational ($s=2$), axial vector ($s=1$), and scalar ($s=0$) tidal perturbation are then defined as
\begin{equation}
\begin{split}
&k^{(s=2)}_{l}\equiv -\frac12 \frac{l(l-1)}{M^{2l+1}} \sqrt{\frac{4\pi}{2l+1}}\frac{M_l}{\mathcal{E}_{l}},\\
& k^{(s=1)}_{l}\equiv -\frac12 \frac{l(l-1)}{M^{2l+1}}\sqrt{\frac{4\pi}{2l+1}}  \frac{J_l}{\mathfrak{B}_{l}}\,, \\
& k^{(s=0)}_{l}\equiv -\frac12 \frac{l(l-1)}{M^{2l+1}}\sqrt{\frac{4\pi}{2l+1}}  \frac{\phi_l}{\mathcal{E}^\text{S}_l}\,.
\label{Lovenumbersdef1}
\end{split}
\end{equation}
Note that for $s=0,1$ the normalization of the above definitions is arbitrary and was chosen to be equivalent to the $s=2$ case.

The TLNs defined above can be computed by solving the corresponding perturbation equations, requiring regularity at the BH horizon, and finally matching the solution with the asymptotic expansion. This is discussed below for the scalar and axial-vector cases, separately.

\subsection{Scalar Love numbers}
The field equation for the external scalar field $\phi$ (not to be confused with the scalar field $\Phi$ that condensates around the BH and that  is part of the background) is given by the (real) Klein-Gordon equation
\be
\Box \phi (v,r, \theta, \varphi) = 0.
\ee
Owing to the spherical symmetry of the background  and focusing on static perturbations (which are the relevant ones to compute the TLNs that enter the GW signal at the leading-order in a PN expansion), it is again convenient to perform a spherical-harmonic decomposition,
\begin{equation}
 \phi = \sum_{lm} \frac{R (r)}{r}  Y^{lm}\,,
\end{equation}
which yields the following radial equation
\be
  F\frac{R''}{r} + \lp  F' + F  \delta \lambda' \rp \frac{R'}{r} - \lp
 \frac{l(l+1)}{r^2} + \frac{1}{r} F' + \frac{F}{r}  \delta \lambda' \rp \frac{R}{r}  = 0\,, \label{eqScalarl}
\ee
where the prime denotes differentiation with respect to $r$.
Since we wish to use this computation as a proxy for the gravitational case (for which $l\geq2$), we shall focus on the $l=2$ scalar mode.

Equation~\eqref{eqScalarl} can be solved perturbatively in the coupling strength of the source, given in terms of a dimensionless parameter $\epsilon = M_\BH^2 \Phi_0^2$.
At zeroth order in $\epsilon$ the solution has the asymptotic form at spatial infinity
\be
\label{leading}
\frac{R^{(0)}}{r} = 6 c_1 \lp \frac{r}{r_\BH} \rp^2,
\ee
after imposing the regularity of the solution at the BH horizon\footnote{Note that, since the background solution is perturbative in the matter field, regularity should be imposed on the horizon of the naked BH, which is also the singular point of Eq.~\eqref{eqScalarl}, both to zeroth and to first order in the matter fields.}. The fact that the solution above does not have an induced quadrupolar term $\sim1/r^3$ implies that the scalar TLNs of a naked Schwarzschild BH are zero~\cite{Cardoso:2017cfl,Hui:2021vcv}.
The solution at first order in $\epsilon$ can be obtained analitically 
and has the following asymptotic form at spatial infinity
\be
\frac{R^{(1)}}{r} = -  \mathcal{N} c_1 \frac{\epsilon}{G^{10} M_\BH^{11} \mu^{12}} \frac{1}{r^3} = -\frac{\mathcal{N}}{8} c_1 \frac{\epsilon}{M_\BH^{14} \mu^{12}} \lp \frac{r_\BH}{r} \rp^3,
\ee
where the numerical coefficient $\mathcal{N}$ is determined  depending on the considered background,
such that the total solution is given by
\be
\frac{R}{r} = \frac{R^{(0)}}{r} +\frac{R^{(1)}}{r} = 6 c_1 \lp \frac{r}{r_\BH} \rp^2 \llp 1 - \frac{\mathcal{N}}{48} \frac{\epsilon}{ M_\BH^{14} \mu^{12}} \lp \frac{r_\BH}{r} \rp^5 \rrp.
\ee
By comparing the asymptotic behavior of the solution to Eq.~\eqref{eq:scalarexpansion}
one can extract the scalar TLN as\footnote{We shall use the notation $k_l^{(\ell,s)}$ to denote the $l$-th order TLN of a tidal field with spin $s$ for a BH dressed with a background scalar condensate with harmonic index $\ell$.}
\be
k_2^{(\ell, s=0)}  = -\frac{\mathcal{N}^{(\ell)}}{96} \frac{\epsilon^{(\ell)}}{M_\BH^{14} \mu^{12}},
\ee
where we have highlighted the dependence of the numerical coefficient and the expansion parameter on the mode $\ell$ of the complex scalar field background $\Phi$. Note that, in the adiabatic regime, the final result does not depend explicitly on $\tilde v$.
The leading order term of the dimensionless TLN is given for the mode $\ell=0,1,2$ modes of the background condensate as
\begin{align}
k_2^{(\ell=0,s=0)} &= -\frac{27}{192} \pi \frac{|\Phi_0^{\ell=0}|^2}{\alpha^{12}} e^{-4\alpha^2}, \nonumber \\
k_2^{(\ell=1,s=0)} &= -1008 \pi \frac{|\Phi_0^{\ell=1}|^2}{\alpha^{12}} e^{-2\alpha^2}, \nonumber \\
k_2^{(\ell=2,s=0)} &= -\frac{6200145}{4} \pi \frac{|\Phi_0^{\ell=2}|^2}{\alpha^{12}} e^{-\frac{4}{3}\alpha^2},
\end{align}
which can be finally expressed in terms of the gravitational coupling $\alpha$ and the scalar cloud mass $M_s$ as
\begin{align}
k_2^{(\ell=0,s=0)} &\simeq -0.14 \frac{1}{\alpha^8} \lp \frac{M_s}{M_\BH} \rp, \nonumber \\
k_2^{(\ell=1,s=0)} &\simeq -10 \frac{1}{\alpha^8} \lp \frac{M_s}{M_\BH} \rp, \nonumber \\
k_2^{(\ell=2,s=0)} & \simeq -160  \frac{1}{\alpha^8} \lp \frac{M_s}{M_\BH} \rp.
\end{align}
One can appreciate that higher modes of the complex scalar background experience larger deformability, 
and hence a larger value of the TLN. This is due to the fact that boson condensates with larger values of $\ell$ are more diluted, and hence more deformable.

We also find, as expected, that the behavior of the Love number in terms of the gravitational coupling $\alpha$ reproduces the one found in Ref.~\cite{Baumann:2018vus},
using a dimensional analysis for a gravitational tidal perturbation.

\subsection{Axial vector Love numbers}
\noindent
The equation of motion for a vector field $A_\mu$ in the BH+cloud spacetime is given by
\be
\nabla_\nu F^{\mu \nu} = 0,
\ee
where we have defined the electromagnetic tensor $F_{\mu \nu} = \partial_\mu A_\nu - \partial_\nu A_\mu$.
For simplicity in this case we shall focus on the axial sector only and decompose the spin-$1$ field as (see, e.g., Ref.~\cite{Pani:2013wsa})
\begin{equation}
 A_\mu = \sum_{lm} \lp 0, 0, a^{lm}(r) \frac{\partial_{\varphi} Y^{lm}}{\sin \theta}, -a^{lm}(r) \sin \theta  \partial_{\theta} Y^{lm}\rp\,.
\label{fieldsansatz}
\end{equation}
Inserting this decomposition in the equation of motion gives
\be
F a'' + \lp  F'+ F  \delta \lambda' \rp a' -
\frac{l(l+1)}{r^2} a  = 0,
\ee
where $a\equiv a^{lm}$. Also this equation can be solved perturbatively in the coupling strength of the background matter source. For $l=2$, after imposing the regularity of the solution at the BH horizon, the solution at zeroth order has the following asymptotic form at spatial infinity
\be
a^{(0)} = c_1 \lp \frac{r}{r_\BH} \rp^3\,.
\ee
Again, the absence of an induced quadrupolar term implies that the axial vector TLNs of a naked Schwarzschild BH are zero.
On the other hand, at first order the solution has the asymptotic form 
\be
a^{(1)}= -  \mathcal{N} c_1 \frac{\epsilon}{M_\BH^{11} \mu^{12}} \frac{1}{r^2} = -\frac{\mathcal{N}}{4} c_1 \frac{\epsilon}{M_\BH^{14} \mu^{12}} \lp \frac{r_\BH}{r} \rp^2,
\ee
such that the total solution is given by
\be
a = a^{(0)} +a^{(1)}= c_1 \lp \frac{r}{r_\BH} \rp^3 \llp 1 - \frac{\mathcal{N}}{4} \frac{\epsilon}{M_\BH^{14} \mu^{12}} \lp \frac{r_\BH}{r} \rp^5 \rrp,
\ee
from which one can extract the axial-vector TLN as
\be
k_2^{(\ell,s=1)}  = -\frac{\mathcal{N}^{(\ell)}}{8} \frac{\epsilon^{(\ell)}}{M_\BH^{14} \mu^{12}}.
\ee
The leading-order term of the dimensionless TLN for the $\ell=0,1,2$ modes of the complex scalar field background $\Phi$ is then given by
\begin{align}
k_2^{(\ell=0, s=1)} &= - \frac{81}{256}\pi \frac{|\Phi_0^{\ell=0}|^2}{\alpha^{12}} e^{-4\alpha^2}, \nonumber \\
k_2^{(\ell=1, s=1)} &= -2268 \pi \frac{|\Phi_0^{\ell=1}|^2}{\alpha^{12}} e^{-2\alpha^2}, \nonumber \\
k_2^{(\ell=2, s=1)} &= -\frac{55801305}{16} \pi \frac{|\Phi_0^{\ell=2}|^2}{\alpha^{12}} e^{-\frac{4}{3}\alpha^2},
\end{align}
which can be expressed as
\begin{align}
k_2^{(\ell=0, s=1)} &\simeq -0.32 \frac{1}{\alpha^8} \lp \frac{M_s}{M_\BH} \rp, \nonumber \\
k_2^{(\ell=1, s=1)} &\simeq -24 \frac{1}{\alpha^8} \lp \frac{M_s}{M_\BH} \rp, \nonumber \\
k_2^{(\ell=2, s=1)} &\simeq -360 \frac{1}{\alpha^8} \lp \frac{M_s}{M_\BH} \rp.
\end{align}
We stress that, as in the case of the scalar TLNs, higher modes of the complex field $\Phi$ are characterized
by a larger response to the tidal perturbation, and that the scaling with the gravitational coupling $\alpha$ matches the one obtained for a scalar or gravitational tidal perturbation. Indeed, we confirm the general scaling predicted in Ref.~\cite{Baumann:2018vus}, namely
\begin{equation}
 k_2 \sim  \frac{1}{\alpha^8} \lp \frac{M_s}{M_\BH} \rp,
\end{equation}
with the prefactor depending on the background solution and on the type of tidal perturbation. Interestingly, the above scaling shows that $k_2$ can be very large in the $\alpha\ll1$ regime.

Finally, adopting the same normalization used in the gravitational case, the axial vector Love numbers are slightly larger than the scalar ones for the same background configuration.
In Appendix~\ref{appendix} we shall compute the scalar, axial vector, and gravitational TLNs for a perfect-fluid neutron star and show that the same hierarchy holds, namely for a given background configuration the TLNs slightly grow with the value $s=0,1,2$ of the spin of the external tidal field.
Based on this (small) hierarchy, using the axial vector TLNs as a proxy for the real gravitational TLNs should provide a conservative estimate, since the gravitational TLNs are expected to be slightly larger. In the following we shall focus our analysis on the axial vector TLNs.

\section{Measuring the tidal deformability of dressed BHs}

\subsection{Data analysis}
\noindent
To estimate the detectability of the TLNs through GW observations we use a Fisher matrix approach as described in Refs.~\cite{Poisson:1995ef, Vallisneri:2007ev, Cardoso:2017cfl}, which we
follow to summarize the basic notions of the formalism.

The output $s(t)$ of a generic interferometer can be expressed as the sum of the GW signal $h (t,\vec \xi)$ and the stationary detector noise $n(t)$. The posterior on the hyperparameters $\vec \xi$ is given by
\be
p (\vec \xi| s) \propto \pi (\vec \xi) e^{-\frac{1}{2}(h (\vec \xi) - s|h (\vec \xi) - s)},
\ee
in terms of the prior $\pi (\vec \xi)$, where we have defined the inner product as
\be
(g|h) = 2\int_{f_\text{\tiny min}}^{f_\text{\tiny max}} \d f \frac{h (f) g^* (f) + h^*(f) g(f)}{S_n(f)},
\ee
where $S_n(f)$ is the detector noise spectral density, and $f_\text{\tiny min}$ and $f_\text{\tiny max}$ are the characteristic minimum and maximum frequencies of integration, which will be discussed in the next section for the system under consideration.

According to the principle of the maximum-likelihood estimator, the values of the hyperparameters can be estimated as those for which the posterior is maximum. In the limit of large signal-to-noise ratio~(SNR$=\sqrt{(h|h)}$), for which the posterior peaks around the true values $\vec \zeta$ of the hyperparameters, one can perform a Taylor expansion to get
\be
p (\vec \xi| s) \propto \pi (\vec \xi) e^{-\frac{1}{2}\Gamma_{ab} \Delta \xi^a \Delta \xi^b},
\ee
with $\Delta \vec \xi = \vec \zeta - \vec \xi$, and where we have introduced the Fisher matrix, 
\be
\Gamma_{ab} = \lp \frac{\partial h}{\partial \xi^a} \bigg | \frac{\partial h}{\partial \xi^b} \rp_{\vec \xi = \vec \zeta}\,.
\ee
The errors on the hyperparameters are given by $\sigma_a = \sqrt{\Sigma^{aa}}$, where $\Sigma^{ab} = \lp \Gamma^{-1}\rp^{ab}$ is the covariance matrix.

We will assume that the GW waveform is approximated by the   sky-averaged, spin-aligned TaylorF2 template in the frequency domain~\cite{Damour:2000gg}
\be
 h (f) = \mathcal{A} f^{-7/6} e^{i \psi (f)},
\ee
where the amplitude is given by~\cite{Berti:2004bd}
\be
\mathcal{A} = \frac{1}{\sqrt{30}\pi^{2/3}} \frac{\mathcal{M}^{5/6}}{D_L},
\ee
in terms of the luminosity distance $D_L$ and chirp mass $\mathcal{M}$ of the binary, while
the dependence on the binary parameters is captured in the GW phase $\psi$. The latter is expressed as the sum of three different contributions in a post-Newtonian~(PN) expansion: the point-particle contribution 
up to 3.5PN order~\cite{Damour:2000gg, Arun:2004hn, Buonanno:2009zt, Abdelsalhin:2018reg}, the tidal heating\footnote{The contribution of the tidal heating depends on the energy absorbed at the horizon and is proportional to the BH cross section. Since the BH radius is affected by the cloud only by a ${\cal O}(\alpha^6)$ correction (see Eqs.~\eqref{rH} and \eqref{Phi0}), we shall assume the same tidal-heating term for naked and dressed BHs to the leading order in $\alpha$. At any rate, the contribution of the tidal heating is typically small and negligible for our analysis.} contribution which introduces a 2.5PN(4PN) $\times \log \beta$ correction to the phase for spinning (nonspinning) binaries relative to the leading term~\cite{Alvi:2001mx,Maselli:2017cmm}, and the tidal deformability contribution $\psi_\text{\tiny TD}(x) = \Lambda  x^5 + \delta\Lambda x^6$, 
where the $5$PN and $6$PN terms are given by \cite{Flanagan:2007ix,Vines:2011ud}
\begin{align}
 \Lambda &=\left(264 -\frac{288}{\eta_1}\right) \frac{\lambda_2^{(1)}}{M^5} +    (1\leftrightarrow 2)\,, \nonumber \\
 \delta\Lambda&=\left(  \frac{4595}{28}- \frac{15895}{28 \eta_1}  + \frac{5715 \eta_1}{14} -
 \frac{325 \eta_1^2}{7}   \right)\frac{\lambda_2^{(1)}}{M^5} +(1\leftrightarrow 2) \,,
\end{align}
in terms of the quadrupolar TLN, $\lambda_2^{(A)} = 2m_{A}^5 k_2^{(A)}/3$, of the $A$-th body.
The PN expansion is given in terms of $x=(M\omega)^{2/3} = \beta^2$, where $\omega$ is the orbital angular velocity,  $\nu = \eta_1 \eta_2$ with $\eta_i = m_i/M$, and $M = m_1 + m_2$.

Note that the TaylorF2 waveform approximant is valid only at small frequencies and becomes increasingly less accurate as the binary approaches the merger. For the system at hand this is not a limitation, because --~due to the possible tidal disruption of the background condensate discussed in the next section~-- the maximum frequency $f_\text{\tiny max}$ is much smaller than the merger frequency.

Given  that the TLNs $k_2$ of the two bodies depend only on their masses and on the coupling $\alpha$, the set of intrinsic hyperparameters can be taken to be $\vec \xi = (\phi_c, t_c, \ln \mathcal{M}, \ln \nu, \ln \alpha, \chi_1, \chi_2)$ (phase and time at the coalescence, chirp mass, symmetric mass ratio, gravitational coupling, and individual spins). A GW measurement of the tidal deformability will automatically translate into a measurement of the coupling $\alpha$ and, hence, on the scalar field mass. 

Given that the spin parameters have values smaller than unity, we have implemented a prior on their magnitude assuming a Gaussian distribution centered around zero with large variance $\sigma_\chi = 2$. This can be achieved by adding to the standard Fisher matrix $\Gamma$ another matrix $\Gamma_\chi = {\rm diag} (0, \dots, 0, 1/\sigma_\chi^2, 1/\sigma_\chi^2)$, following Ref.~\cite{Poisson:1995ef}.

Finally, we have assumed the analytic fits for the noise spectral density provided in Ref.~\cite{Sathyaprakash:2009xs} for ET and the most optimistic LISA configuration 
with $5 \times 10^6 \, {\rm km}$ arm-length and an observing time of $T_\text{\tiny obs} = 5 \, {\rm yr}$ given in Ref.~\cite{Klein:2015hvg}. We also stress that for LISA the amplitude of the GW waveform has been multiplied by an additional factor $\sqrt{3}/2$, in order to account for the triangular shape of the detector~\cite{Cutler:1997ta, Berti:2005ys}, and by a factor $\sqrt{3/20}$ for a proper inclusion of a sky average~\cite{Berti:2004bd} as done in Ref.~\cite{Cardoso:2017cfl}.

\subsection{Relevant frequencies}
Compared to other standard inspiral analyses, for the problem at hand the frequency domain of the Fisher analysis is particularly relevant because it is dictated by some physical requirements.

The maximum frequency, $f_\text{\tiny max}$, is obtained as the minimum between three characteristic frequencies:
\begin{enumerate}
\item[(i)] the first frequency is determined by the fact that, as the inspiral proceeds, tidal forces may disrupt the scalar condensate around the BH. This happens when the semi-major axis of the binary is of the order of the Roche radius~\cite{Shapiro:1983du} 
	\begin{align}
	r_\text{\tiny Roche} &= \gamma r_\text{\tiny BH} \lp \frac{\rho_\text{\tiny BH}}{\rho} \rp^{1/3} = \gamma  r_s \lp \frac{M_\text{\tiny BH}}{M_s} \rp^{1/3} \nonumber \\
	&= \frac{\gamma(\ell+1)^2}{2} \frac{r_\BH}{\alpha^2}    \lp \frac{M_\text{\tiny BH}}{M_s} \rp^{1/3},
	\end{align}
	where the numerical coefficient $\gamma \sim \mathcal{O} (2)$ takes values from 1.26 for  rigid bodies to 2.44 for fluid ones. We stress that we have assumed the primary object to be dominated by the BH itself (neglecting its halo) with density $\rho_\text{\tiny BH} = 3M_\BH/4 \pi r^3_\BH$, destroying the halo of the secondary with density $\rho = 3M_s/4 \pi r^3_s$.  
	This is consistent since in our perturbative scheme $\rho\ll \rho_\BH$, as previously discussed.
	The corresponding GW frequency for a circular, equal-mass binary is given by 
	\be
	f_\text{\tiny Roche} = \frac{2^{3/2}}{(\ell+1)^{3} \pi \gamma^{3/2}} \frac{\alpha^3}{r_\BH} \lp \frac{M_s}{M_\text{\tiny BH}} \rp^{1/2};
	\ee
\item[(ii)] the second frequency is determined by the applicability of the multipole expansion in the PN waveform, which holds for semi-major axis larger than the cloud radius. This translates into 
	\be
	r \gtrsim \frac{(\ell+1)^2}{2} \frac{r_\BH}{\alpha^2},
	\ee
	or into a maximum frequency
	\be
	f_\text{\tiny me} = \frac{2^{3/2}}{(\ell+1)^3 \pi}\frac{\alpha^3}{r_\BH};
	\ee
\item[(iii)] the third frequency is the standard cutoff of the early-inspiral waveform approximants, which is approximately set at the frequency at the innermost circular orbit radius~(ISCO),
	\be
	f_\text{\tiny ISCO} = \frac{1}{6\sqrt{6} \pi r_\BH}\,.
	\ee
\end{enumerate}
Therefore, we set $f_\text{\tiny max}  = {\rm Min} \llp f_\text{\tiny Roche},  f_\text{\tiny me}, f_\text{\tiny ISCO}\rrp$. It is easy to see that, for the scenarios we are interested in, the smallest frequency is always the Roche one. That is, well before the binary approaches the ISCO, the tidal field can destroy the condensate. Therefore for $f>f_\text{\tiny Roche}$ the tidal effects are negligible and the binary proceeds until the merger of two naked BHs. 
Note that  larger values of the condensate multipole $\ell$ would result in a smaller maximum frequency.
Likewise, $f_\text{\tiny Roche}\propto\alpha^3/M_\BH$. This strong dependence on $\alpha$ severely limits the detectability of tidal effects and is competitive with the fact that $k_2\propto \alpha^{-8}$. Despite the fact that $k_2$ grows very rapidly as $\alpha\to0$, one can easily check that the PN correction $\Lambda x^{10}$ is much smaller than unity at $f=f_\text{\tiny Roche}$, so the PN expansion is valid in the relevant frequency range.

\begin{figure}[t!]
	\centering
	\includegraphics[width=1\linewidth]{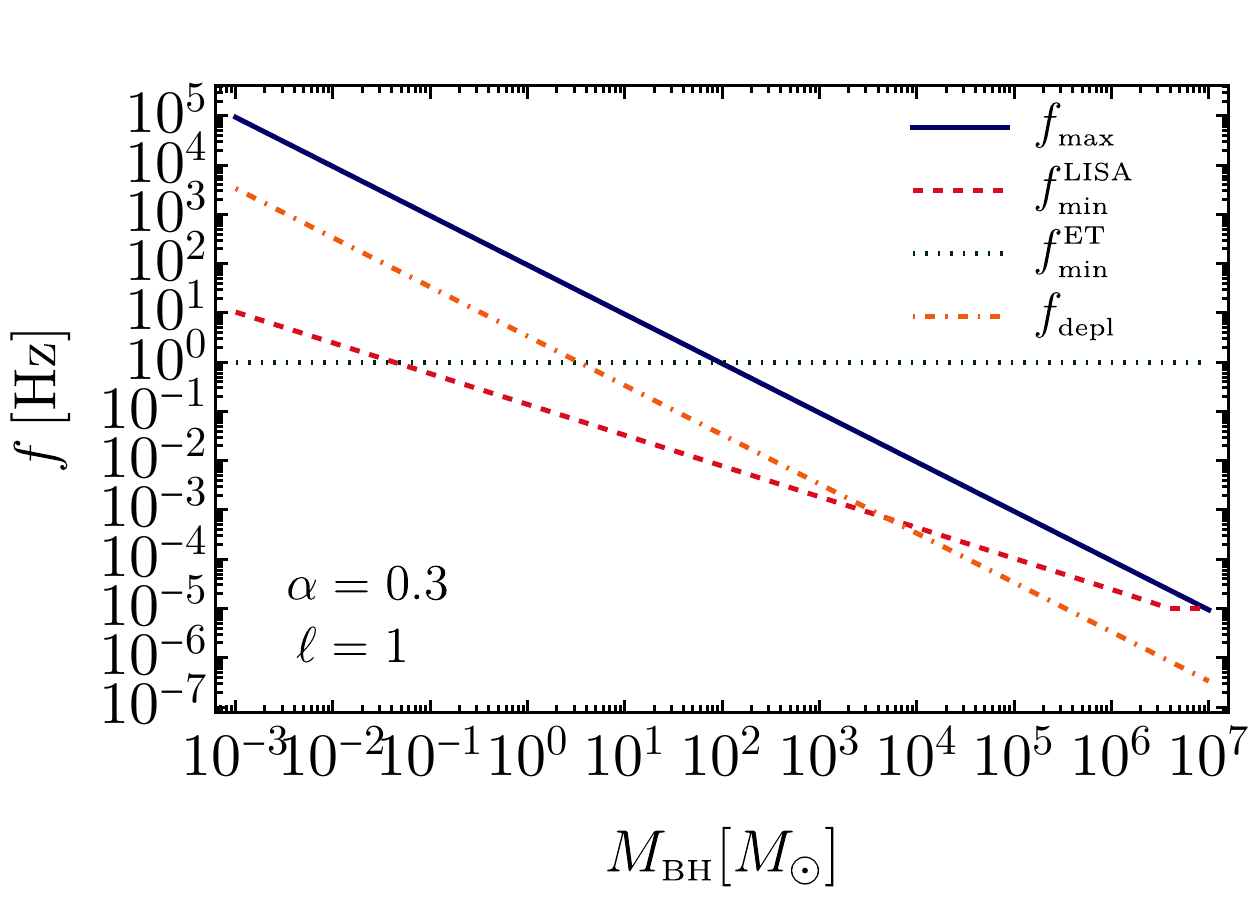}
	\caption{\it Comparison between the relevant frequencies in terms of the mass of a equal-mass binary system, for a fixed value of the gravitational coupling $\alpha$ and for the mode $\ell = 1$ of the scalar field background.}
	\label{fig:fre}
\end{figure}

Let us now move to the choice of the minimum frequency, $f_\text{\tiny min}$.
In the standard (naked) BH case one typically sets $f_\text{\tiny min}$ to the minimum frequency detectable by the interferometer. For ET we consider $f^\text{\tiny ET}_\text{\tiny min} = 1\,  {\rm Hz}$, whereas for LISA we take  $f^\text{\tiny LISA}_\text{\tiny min} = {\rm Max}[10^{-5} \, {\rm Hz}, f_\text{\tiny 5yr}$],  i.e. either $10^{-5} \, {\rm Hz}$ or the initial frequency corresponding to a binary that spends $T_\text{\tiny obs}=5\,{\rm yr}$ to span the frequency band up to $f_\text{\tiny max}$~\cite{Berti:2004bd}
\be
f_\text{\tiny 5yr} = \llp \frac{1}{f_\text{\tiny max}^{8/3}} + 1.27 \times 10^{12} \, {\rm Hz} \lp \frac{\mathcal{M}}{10^6 M_\odot} \rp^{5/3} \lp \frac{T_\text{\tiny obs}}{5 \rm yr} \rp \rrp^{-3/8}.
\ee
However, in the case of BHs surrounded by superradiantly-triggered bosonic condensates one has also to consider the gravitational perturbation due to the presence of the BH companion, which induces a transition of the dominant growing mode of the scalar condensate to decaying modes, resulting into  a depletion of the cloud. 
For nonaxisymmetric condensates (as those produced by the BH superradiant instability), this condition translates into the  requirement that  the time to merger has to be shorter than the lifetime of the cloud, which holds for frequencies larger than the critical value~\cite{Baumann:2018vus} (see also Refs.~\cite{Berti:2019wnn,Cardoso:2020hca,Takahashi:2021eso} for recent studies on the impact of depletion on axion clouds)
\begin{equation}
 f_\text{\tiny depl} = 0.3 \,{\rm mHz} \lp \frac{3 M_\odot}{M_\BH} \rp \frac{(1+q)^{1/8}}{q^{3/8}} \lp \frac{\alpha}{0.07} \rp^{45/8}\,,
\end{equation}
where $q$ is the binary's mass ratio.
This frequency is found to be smaller than the maximum frequency dominated by the Roche value for the relevant superradiant modes $\ell = 1,2$, and is competitive with the sensitivity frequencies of the considered experiments in setting the minimum frequency. 
In practice, to account for this effect, when dealing with $\ell=1,2$ condensates we set $f_\text{\tiny min}={\rm Max}[f_\text{\tiny min}^\text{\tiny detector}, f_\text{\tiny depl}]$, whereas for spherical ($\ell=0$) condensates there are no level transitions and we simply set $f_\text{\tiny min}=f_\text{\tiny min}^\text{\tiny detector}$.
 The comparison between the relevant frequencies is shown in Fig.~\ref{fig:fre} for a fixed value of the gravitational coupling and the mode $\ell = 1$ of the scalar condensate.

\begin{figure*}[t!]
	\centering
	\includegraphics[width=0.48 \linewidth]{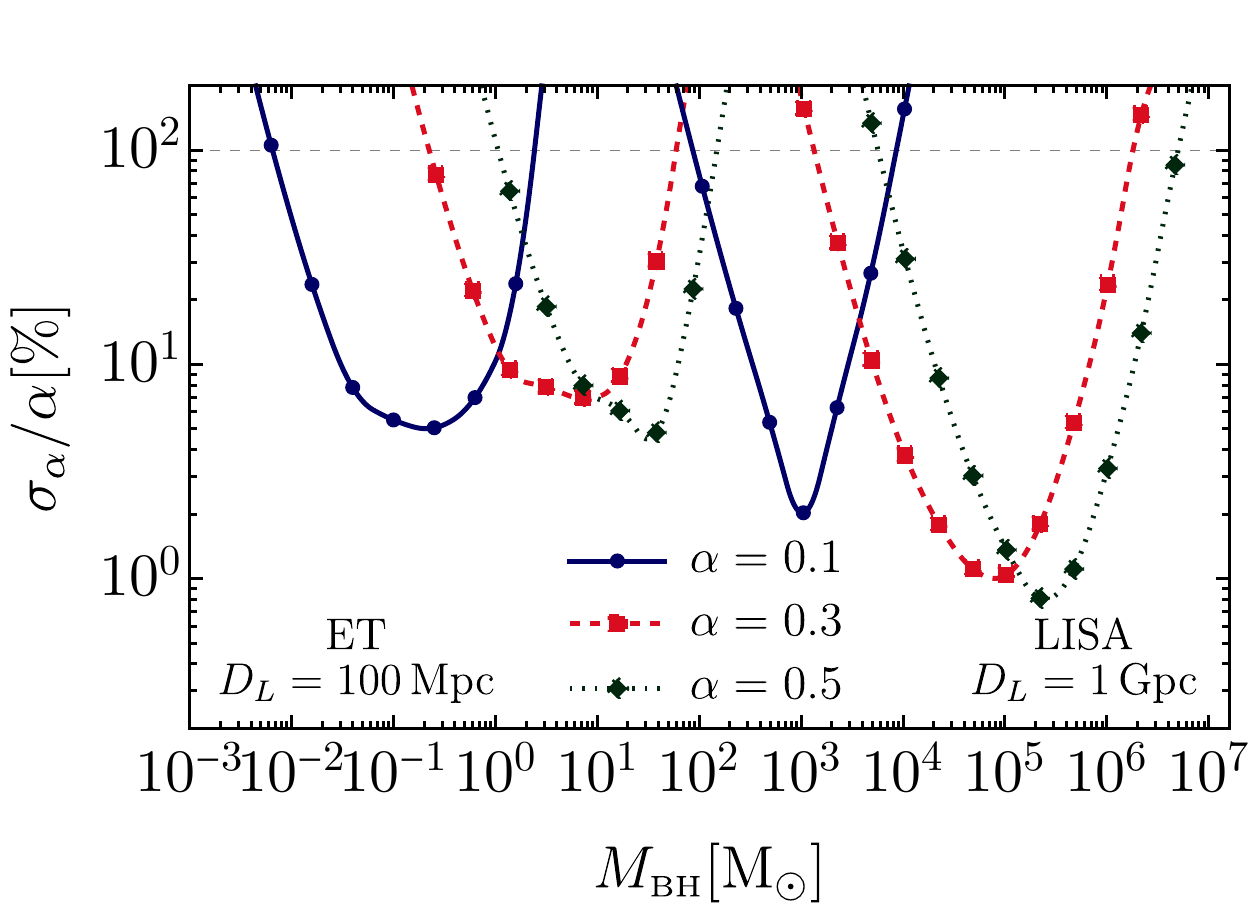}
	\includegraphics[width=0.48 \linewidth]{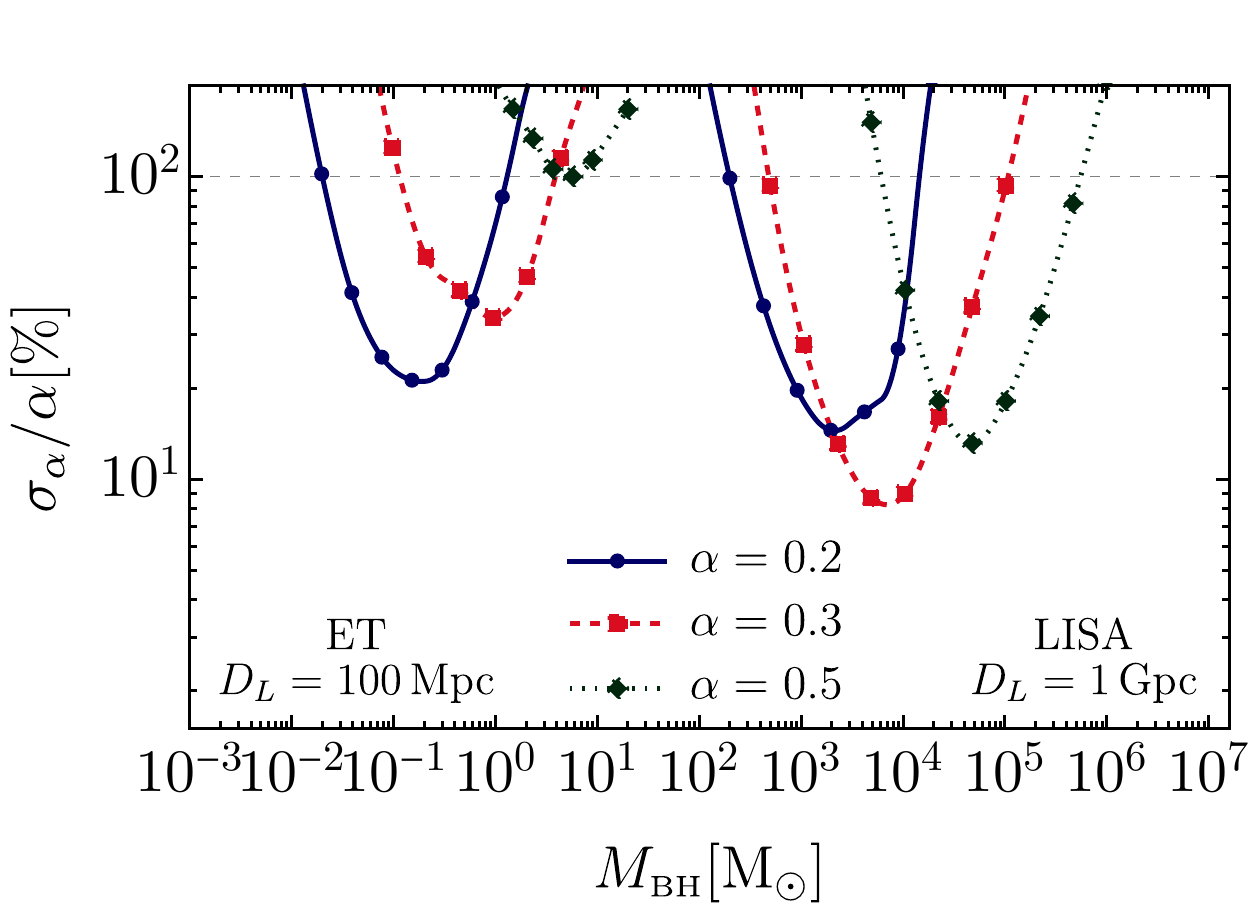}
	\caption{\it Relative percentage error on the coupling $\alpha=\mu M_\BH$ for equal-mass, nonspinning binaries for the modes $\ell = 0$ (left) and $\ell=1$ (right) of the scalar field condensate, assuming an axial vector TLN in the waveform and $M_s (\alpha) = 0.1 M_\BH$. 
	The left-most (right-most) curves refer to binaries detected by ET (LISA) at luminosity distance $D_L=100\,{\rm Mpc}$ ($D_L=1\,{\rm Gpc}$).
	The horizontal dashed lines identify the value $\sigma_\alpha/\alpha = 1$ above which no constraint can be set on the gravitational coupling.  
	}
	\label{fig:2}
\end{figure*}
\subsection{Numerical results}
Our main numerical results are summarized in Fig.~\ref{fig:2}, where we show the relative error on the gravitational coupling $\alpha$ for a nonspinning, equal-mass binary, for the $\ell=0$ and $\ell=1$ modes of the complex scalar field $\Phi$. 
We considered values of $\alpha={\cal O}(0.1)$ for which the analytical result valid in the Newtonian ($\alpha\ll1$) limit is a reasonable approximation~\cite{Yoshino:2013ofa}.
In all cases, for the computation of the Love numbers we considered $M_s=0.1 M_\BH$, which corresponds to the most optimistic scenario, since smaller values of $M_s$ makes the TLN smaller and, most importantly, reduce the Roche frequency.

The main finding is that both ET and LISA would be able to constrain the gravitational coupling, and thus the tidal deformability, in a wide and complementary range of BH masses. 
Remarkably, in both cases the bounds extend to smaller BH masses than those typically probed by inspiral tests. The reason is that the characteristic GW frequency of inspiral tests is roughly $f_\text{\tiny ISCO}\approx 0.01/M_\BH$, which matches the optimal frequency of ground-based (space-based) detectors --~i.e., roughly $100\,{\rm Hz}$ ($1\,{\rm mHz}$)~-- for $M_\BH\sim 20\,M_\odot$ ($M_\BH\sim 2\times 10^6\,M_\odot$). On the other hand, the Roche frequency is roughly 
\begin{equation}
 f_\text{\tiny Roche} \sim  \frac{5\times 10^{-5}}{(\ell+1)^3 M_\BH}\left(\frac{\alpha}{0.1}\right)^3\left(\frac{M_s}{0.1 M_\BH}\right)^{1/2}\,,
\end{equation}
and, for $\ell=1$, $\alpha=0.1$, $M_s=0.1M_\BH$, matches the optimal frequency of ground-based (space-based) detectors for much smaller BH masses, namely $M_\BH\approx 0.02 M_\odot$ ($M_\BH\approx 2\times10^3 M_\odot$).
Note that, although these frequencies are much smaller than the typical merger frequency, tidal effects are still detectable because of the large magnitude of the TLNs in the small-$\alpha$ limit.
For this reason an experiment like LISA could detect the tidal deformability of dressed BHs across the entire mass range from stellar-mass ($\approx 10^2 M_\odot$) to supermassive ($\approx 10^7 M_\odot$) objects, providing a measurement of $\alpha$ as accurate as a few percent for optimal configurations. 
It is also interesting to note that the bounds coming from ET and LISA are very well complementary to each other, leaving no gap for intermediate-mass BHs.
In particular, LISA would be sensitive also to tidal-deformability effects in stellar-origin BH binaries well before their merger. The results of Fig.~\ref{fig:2} show that for a GW190521-like~\cite{Abbott:2020tfl} event (with total mass $M\approx 150 M_\odot$ and luminosity distance $D_L\approx5\,{\rm Gpc}$), a coupling $\alpha\approx 0.08$ could be measured with ${\cal O}(50\%)$ accuracy.
Since the effect of the tidal deformability is relevant for this system, not accounting for it could jeopardize multiband detections of stellar-origin BH binaries~\cite{Sesana:2016ljz}, similarly to other environmental effects~\cite{Caputo:2020irr,Toubiana:2020drf}.

By comparing the left and right panels of Fig.~\ref{fig:2} one can notice that smaller-$\ell$ condensates can be better constrained with respect to the higher-$\ell$ configurations. 
The reason for this can be again found by looking at the characteristic maximum frequency of integration in the Fisher analysis. In particular, even though the TLNs for the $\ell=1$ condensate are larger than those for $\ell=0$ by approximately an order of magnitude, the maximum frequency is significantly smaller (given the behavior of $f_\text{\tiny max} \sim (\ell+1)^{-3}$), limiting the capability of the experiment to set a strong bound on the tidal-deformability term that enters at high-PN order.
Indeed, we do not show the bounds for $\ell=2$ condensates since they are not constraining ($\sigma_\alpha/\alpha>1$).

One can also appreciate that for a fixed value of $\alpha$, the relative errors decrease for larger masses, reach a minimum, and then increase again.
The behavior of the curves can be understood as follows: given that the tidal deformability grows with the object mass, higher masses could be easier detected. However, increasing the BH mass would also reduce the number of effective cycles 
in the detector bandwidth due to the maximum frequency dependence on the BH mass $f_\text{\tiny max} \sim 1/M_\BH$ (tidal effects start to get relevant at high PN/high frequency order).
The reason for which curves at different $\alpha$ overlap can again be explained by the strong dependence of the maximum frequency on the gravitational coupling $f_\text{\tiny max} \sim \alpha^{3}$, meaning that larger values of $\alpha$, corresponding to smaller TLNs, would imply a larger number of effective cycles, increasing the possibility for its detection. This is manifestly shown in the plots, where one can see how smaller values of $\alpha$ are better constrained at smaller masses than at larger masses.

\subsection{Projected bounds on ultralight fields from tidal-deformability measurements}

\begin{figure*}[t!]
	\centering
	\includegraphics[width=0.48 \linewidth]{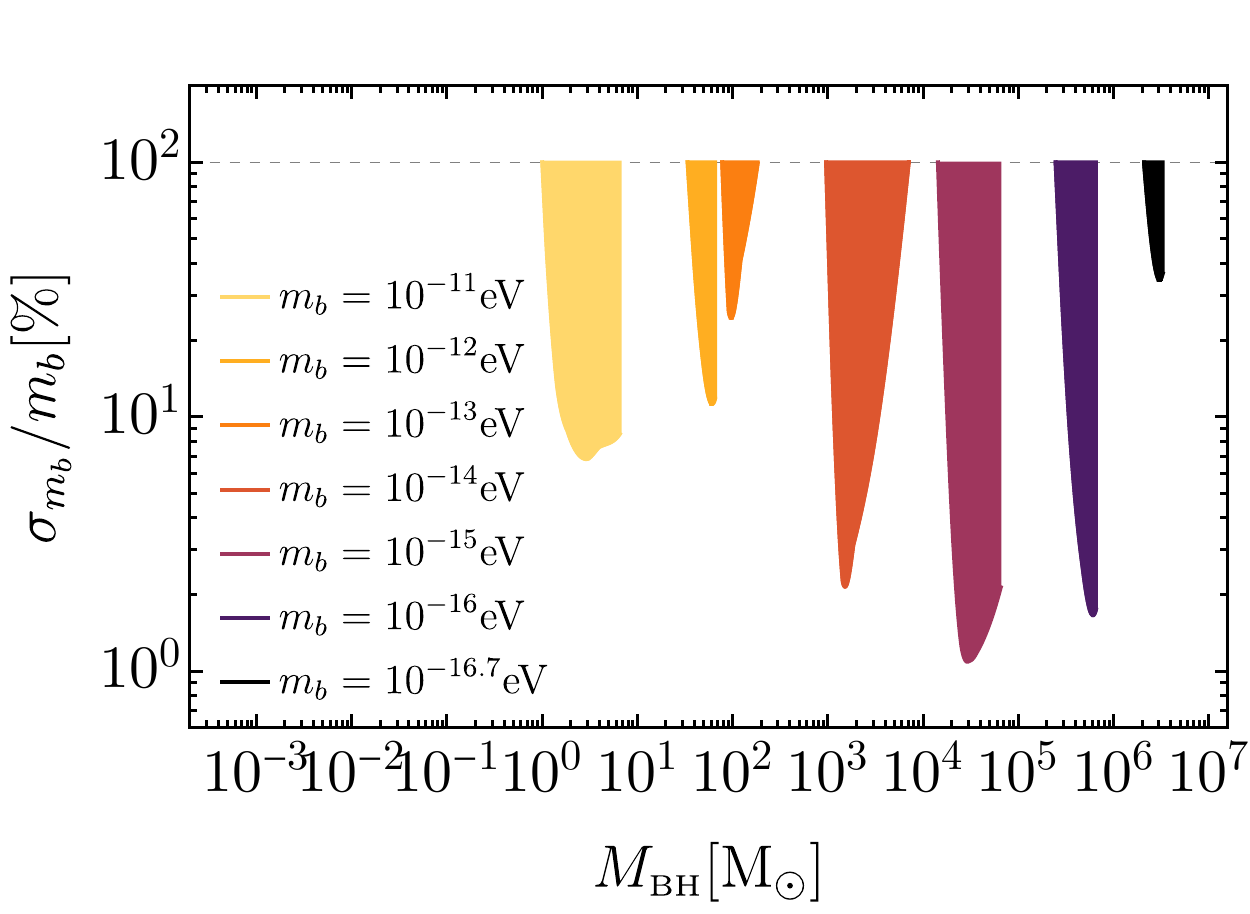}
	\includegraphics[width=0.48 \linewidth]{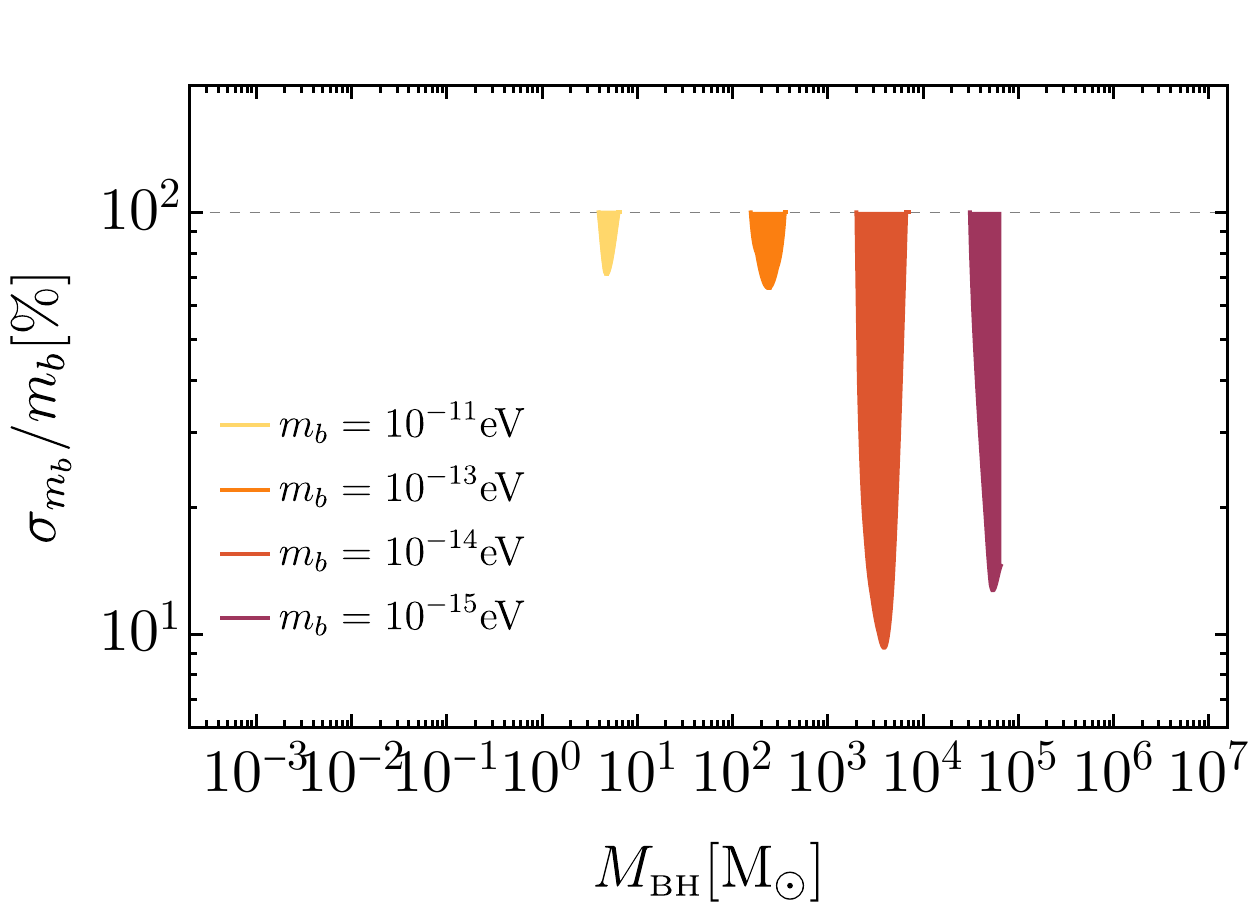}
	\caption{\it Relative percentage error on the scalar field mass  $m_b=\mu \hbar$ for equal-mass, nonspinning binaries for the modes $\ell = 0$ (left) and $\ell=1$ (right) of the condensate, assuming an axial vector TLN in the waveform and $M_s (\alpha) = 0.1 M_\BH$. 
	The horizontal dashed lines identify the value  $\sigma_{m_b}/m_b = 1$  above which no constraint can be set on the scalar field mass.
	}
	\label{fig:3}
\end{figure*}

A measurement/bound on the gravitational coupling $\alpha=M_\BH \mu$ can be immediately translated into a measurement/bound on the boson mass $m_b=\mu \hbar$. The typical SNRs for the systems shown in Fig.~\ref{fig:2} are high enough that the binary chirp mass is always very well measured, also because we restricted to the early inspiral phase to avoid tidal disruption of the condensate.
The binary mass ratio is also measured at least with the same accuracy as $\alpha$, and typically better.
From the measurement errors on the coupling $\alpha$, chirp mass ${\cal M}$, and mass ratio $q$, one can use the standard propagation of errors to determine the relative error on $m_b$, finding that they are very similar to those on $\alpha$, since the  errors on the individual masses are much smaller or at least comparable.
The results shown in Fig.~\ref{fig:2} suggest that the boson mass can be measured with roughly $10\%$ accuracy and down to $1\%$ accuracy in some optimal configurations.

For a given experiment (either ET or LISA) the bounds on the boson mass extend across a certain range.
Since $\alpha={\cal O}(0.1)$ in the relevant parameter space, the lower (upper) boundary of this range is set by the heaviest (lightest) BH that gives a meaningful constraint, i.e. such that $\sigma_\alpha/\alpha<1$.

For spherical ($\ell=0$) condensates, ET can probe values of the coupling slightly smaller than $\alpha=0.1$ down to $M_\BH=10^{-3}M_\odot$, but this of course requires the existence of such light BHs that can only be of primordial origin (see~\cite{Carr:2020gox} for a review on the constraints on primordial BHs). More conservatively, assuming a detection of stellar-origin BHs with $M_\BH\gtrsim 5 M_\odot$, a measurement of $\alpha\approx 0.3$ would translate into a measurement $m_b\approx 10^{-11}\,{\rm eV}$. The upper bound of the range is achieved for $\alpha\approx 0.5$ and $M_\BH\approx 200M_\odot$, corresponding to $m_b\approx 3\times 10^{-13}\,{\rm eV}$.
Overall, ET could probe the range $10^{-13}\lesssim m_b/{\rm eV}\lesssim 10^{-11}$ for $\ell=0$ condensates and a narrower range (especially at high BH masses) for $\ell=1$ condensates (see right panel of Fig.~\ref{fig:2}).
Interestingly, as shown in the left panel of Fig.~\ref{fig:2}, LISA can probe ultralight spherical bosonic condensates in a range that already overlaps with that of ET. Detecting a $100 M_\odot$ BH provides a measurement of $\alpha\approx 0.1$, corresponding to $m_b\approx 10^{-13}\,{\rm eV}$.
In particular, LISA could measure the mass of ultralight bosons with $m_b\approx 10^{-14}\,{\rm eV}$ with $10\%$ accuracy or better. It is interesting to note that the regime around $m_b\approx 10^{-14}\,{\rm eV}$ is particularly difficult to probe with other GW-based tests, see e.g. Refs.~\cite{Hannuksela:2018izj,Yuan:2021ebu} for tests that can marginally probe this range, and Ref.~\cite{Brito:2017zvb} for bounds coming from mass-spin measurements of intermediate-mass BHs with LISA, which are anyway indirect. Therefore, LISA measurements of the tidal deformability of light BHs would be complementary to other searches and would also provide a smoking gun for departure from the standard ``naked BH'' picture.
Overall, LISA could probe the range $10^{-17}\lesssim m_b/{\rm eV}\lesssim 10^{-13}$ for $\ell=0$ condensates, which again becomes slighly narrower for $\ell=1$ condensates.
The LISA range is wider than that of ET both because of the larger SNR and, most importantly, because we expect to detect BHs in the entire mass range available for this measurement.

This discussion is summarized in Fig.~\ref{fig:3}, where we show the relative error on the scalar field mass $m_b$, spanning the range $10^{-17}\lesssim m_b/{\rm eV}\lesssim 10^{-11}$, for both the $\ell=0$ (left panel) and $\ell=1$ (right panel) modes, obtained from the bounds on $\alpha$. In particular, for a fixed value of $m_b$, we perform the Fisher analysis for different values of the BH mass such that $\alpha <0.5$ and translate the bounds on $\alpha$ to bounds on the scalar field mass. One can appreciate that ET and LISA will be able to constrain this range of masses with an accuracy down to few percents, also filling the gap around $m_b \sim 10^{-14} {\rm eV}$ from independent searches.

\section{Conclusions}
In this paper we have provided the computation of the TLNs of a BH surrounded by a bosonic scalar-field condensate under the tidal effects due to scalar and vector perturbations. Our findings show that the system gets deformed, with the TLNs being proportional to the inverse of the eighth power of the gravitational coupling $\alpha=M_\BH \mu$. In particular, small masses of the scalar condensate, $m_b=\mu\hbar\ll 1/M_\BH$, yield very large TLNs, enhancing the effect of the tidal deformability in the waveform. This result is found to be independent on the tidal source, with the ``susceptibility'' of the system to a vector perturbation slightly larger than the one for the scalar tides.

Using the result for the vector tidal perturbation as a proxy for the gravitational TLNs, we have addressed the detectability of tidal effects of dressed BHs with future experiments like the ET and LISA through a Fisher matrix analysis. 
Even focusing only on binaries for which the condensate is surely not destroyed (either by depletion due to mode transitions at small orbital frequency or by tidal disruption at large orbital frequency), we showed that when $\alpha\approx 0.1$ the TLNs are so large that their effects are measurable across various mass ranges.

We showed that LISA could measure the tidal deformability of dressed BHs across the mass range from stellar-mass ($\approx 10^2 M_\odot$) to supermassive ($\approx 10^7 M_\odot$) objects, overall providing a measurement of the mass of ultralight bosons in the range $(10^{-17} - 10^{-13}) \, {\rm eV}$ with roughly $10\%$ accuracy (down to $1\%$ accuracy in some optimal configurations). Interestingly, these constraints extend well within the $10^{-14}\,{\rm eV}$ regime which is particularly hard to probe, thus filling the gap between other superradiance-driven constraints coming from terrestrial and space interferometers. Altogether, our analysis suggests that LISA and ET can probe tidal effects from BHs dressed by scalar condensate in the combined mass range $(10^{-17} - 10^{-11}) \, {\rm eV}$.
 Furthermore, due to the large tidal deformability of bosonic condensates, the TLNs yield a significant contribution also in the relatively early inspiral, thus potentially affecting the dynamics of stellar-origin binaries in the LISA band. If not properly accounted for, this effect could jeopardize multiband detections of these binaries with space- and ground-based detectors.

Our work is intended to be only a first step in understanding the tidal deformability of dressed BHs; as such, it can be extended in several ways: first of all, one should determine the response of the BH-boson condensate to gravitational perturbations and determine its detectability in comparison to the result obtained assuming the vector result as a proxy. This extension should be relatively straightforward for spherically symmetric clouds, whereas an accurate computation of superradiantly-driven bosonic condensates is much more challenging, due to the fact that the condensate is nonaxisymmetric (and time dependent). It is also likely that computing the gravitational TLNs would require to compute the background metric beyond the perturbative approximation adopted in this work.

It would be interesting to determine how the system reacts in the spinning case (see, e.g., Refs.~\cite{Pani:2015nua,Landry:2015cva,Landry:2015zfa,Landry:2017piv,Gagnon-Bischoff:2017tnz,Abdelsalhin:2018reg,Castro:2021wyc} for the case of the tidal deformability of slowly-spinning stars), and estimate whether the spin could impact the bounds on the boson mass.

Finally, we focused on the interesting case of bosonic condensates around BHs, but in principle any matter configuration around BHs (e.g., accretion disks) could feature a nonvanishing tidal deformability~\cite{Cardoso:2019upw}. Quantifying this effect for coalescence GW signals is another interesting extension of this work.

\begin{acknowledgments}
We thank Gabriele Franciolini and Antonio Riotto for insightful discussions and comments on the draft.
V.DL. is supported by the Swiss National Science Foundation (SNSF), project {\sl The Non-Gaussian Universe and Cosmological Symmetries}, project number: 200020-178787. P.P. acknowledges financial support provided under the European Union's H2020 ERC, Starting 
Grant agreement no.~DarkGRA--757480, and under the MIUR PRIN and FARE programmes (GW-NEXT, CUP:~B84I20000100001), and support from the Amaldi Research Center funded by the MIUR program `Dipartimento di Eccellenza" (CUP:~B81I18001170001).
\end{acknowledgments}

\appendix

\section{Scalar, vector, and gravitational TLNs of a neutron star} \label{appendix}
In this appendix we compute the TLNs of  an ordinary (i.e., ``naked'') neutron star under the presence of scalar $(s=0)$ and vector $(s=1)$ perturbations as defined in the main text, and compare the results with the one obtained for a gravitational perturbation $(s=2)$ by Ref.~\cite{Hinderer:2007mb}.

\subsection{Scalar TLNs}
Let us consider the exterior geometry of a spherical static neutron star of mass $M$ and radius $R$ described by the line element
\be
\d s^2 =  - f \d t^2 + f^{-1} \d r^2 + r^2 \d \Omega^2, \qquad f (r) = 1 - \frac{2  M}{r}.
\ee
The equation of motion for a scalar perturbation $\phi$ on this background is given by the real Klein-Gordon equation $\Box \phi  = 0$ which, making use of the spherical-harmonic decomposition shown in the main text, becomes
\begin{align}
  f\frac{R''}{r} +  f' \frac{R'}{r} - \lp
 \frac{l(l+1)}{r^2} + \frac{1}{r} f' \rp \frac{R}{r}  = 0,
\end{align}
where the prime denotes differentiation with respect to $r$.
For the $l=2$ mode, the exterior solution to this equation has the form
\begin{align}
&R_\text{\tiny ext} (r) = c_1 \frac{r \lp 2 M^2 - 6 M r + 3 r^2 \rp}{2 M^2} \nonumber \\
&+ c_2 \llp \frac{3r}{2} - \frac{3 r^2}{2 M} + \frac{r\lp 2 M^2 - 6 M r + 3 r^2 \rp \log \lp \frac{r}{r-2 M} \rp}{4 M^2} \rrp,
\end{align}
whose behavior at asymptotic infinity $r \to \infty$ is given by
\be
\frac{R_\text{\tiny ext} (r)}{r} \to  c_1 \frac{3}{2} \lp \frac{r}{M} \rp^2 \llp 1+ \frac{c_2}{c_1} \frac{4}{45} \lp \frac{M}{r} \rp^5 \rrp.
\ee
By comparing the asymptotic behavior of the scalar perturbation at spatial infinity with Eq.~\eqref{eq:scalarexpansion}, one can extract the TLN,
$k_2 =2c_2/45c_1 \lp M/R \rp^5$, 
which can be expressed  in terms of the star compactness $C = M/R$ and the parameter $y = R \phi_\text{\tiny ext}' (R)/\phi_\text{\tiny ext} (R)$ evaluated at the radius of the neutron star as
\begin{align}
\label{STLNNS}
& k_2^{(s=0)} (y, C) = 4 C^5 (-1+2C) [3(y-2) + 2C (3+y(C-3))] \nonumber \\
& \times \left\{
	-90 C [3 (y-2) + C (12-9y + C (6y-2))]  \right. \nonumber \\
	&\left. + 45 (2C-1) [3(y-2) + 2C (3 + y(C-3))]\log(1-2C)
	 \right\}^{-1}.
\end{align}
The parameter $y$ can be computed by determining the solution of the scalar perturbation $R_\text{\tiny int} (r)$ in the  interior of the neutron star and matching the interior solution to the exterior one at the neutron star radius.
The interior background is described by the metric
\be
\d s^2 = -e^{\nu (r)} \d t^2 + e^{\lambda (r)} \d r^2 + r^2 \d \Omega^2\,,
\ee
with $e^{\lambda (r)} =\lp 1 - 2 m(r)/r \rp^{-1}$. Assuming a perfect-fluid energy momentum tensor,
\be
T_{ab} = (\rho + p) u_a u_b + p g_{ab},
\ee
in terms of the fluid pressure $p$  and density $\rho$,  and the 4-velocity $u_a = (e^{-\nu},0,0,0)$,
one obtains the Tolman-Oppenheimer-Volkoff (TOV) equations of stellar equilibrium
\begin{align}
\frac{\d m}{\d r} &= 4 \pi r^2 \rho, \nonumber \\
\frac{\d p}{\d r} &= - (\rho + p) \frac{m + 4 \pi r^3 p}{r^2 - 2 m r}, \nonumber \\
\frac{\d \nu}{\d r} &= \frac{2(m + 4 \pi r^3 p)}{r^2 - 2 m r},
\end{align}
which can be solved numerically along with the Klein-Gordon equation for the scalar perturbation $\phi$ in the interior background of the neutron star,
\begin{align}
	&\lp 1- \frac{2 m}{r}\rp\frac{R''}{r} + \llp \frac{2 m}{r^2} + 4 \pi r (p-\rho) \rrp\frac{R'}{r} \nonumber \\
	& -  \llp \frac{2 m}{r^3}  + 4 \pi (p-\rho) + \frac{6}{r^2} \rrp \frac{R}{r}  = 0.
\end{align}
In the limit of small compactness (i.e., in the Newtonian limit) one finds
\be
k_2^{(s=0)} \to \frac{2-y}{6+2y} \qquad \text{when} \qquad C \to 0.
\ee
In this limit the interior solution of the scalar perturbation goes as $\phi_\text{\tiny int} \sim c_1 r^2 + c_2 r^{-3}$, 
for which regularity at the origin implies $c_2 = 0$. This will result into $y \to 2$, for which $k \to 0$ as $C \to 0$. 
\begin{figure}[t!]
	\centering
	\includegraphics[width=0.98 \linewidth]{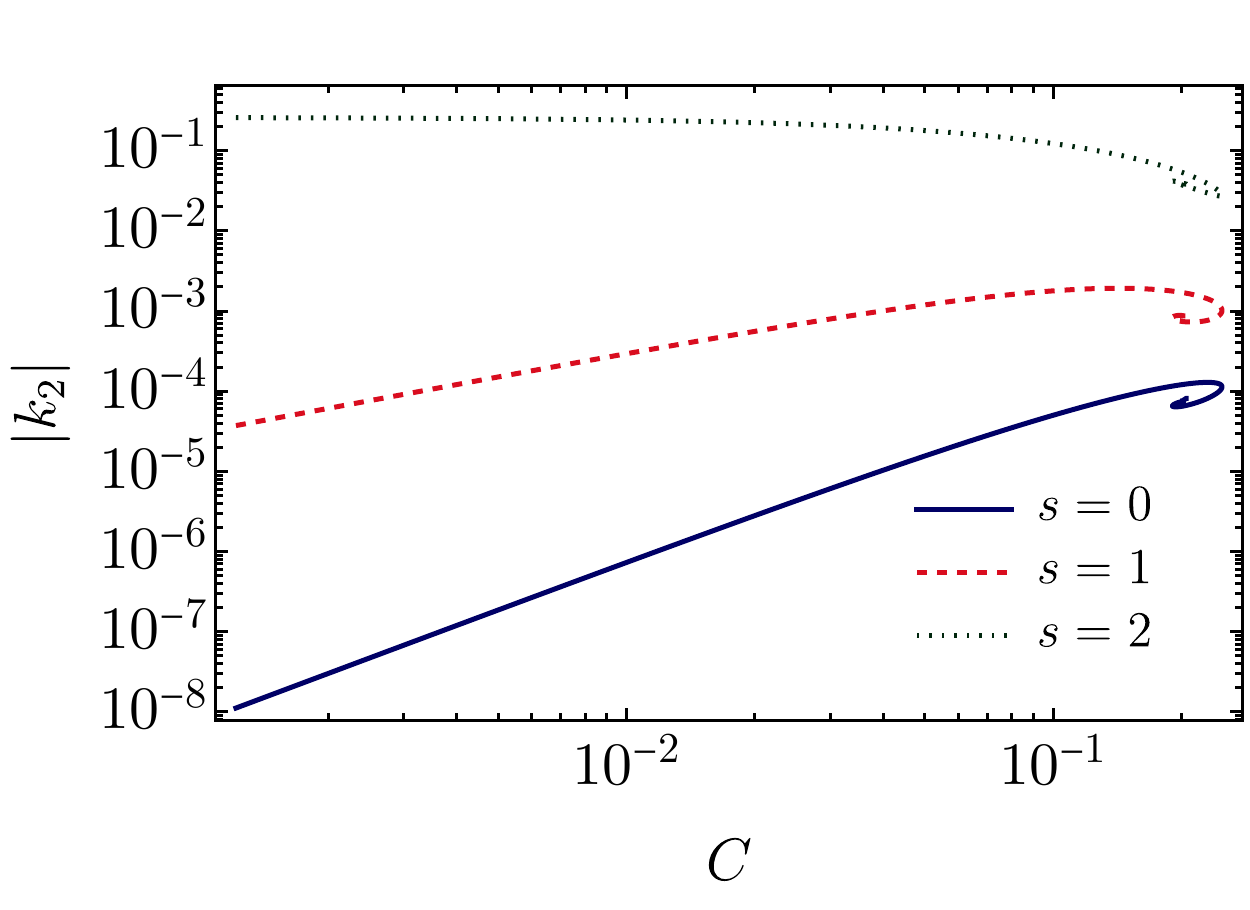}
	\caption{\it Comparison of the $l=2$ TLNs of a neutron star in terms of the compactness for a scalar ($s=0$), vector ($s=1$), and gravitational ($s=2$) tidal perturbation.
	We have assumed the polytropic equation of state~\eqref{EOS} with exponent $\Gamma = 2$ and constant $K = 1/10$. Note that the gravitational TLN is positive, whereas the scalar and axial vector TLNs are negative, as in the BH-condensate case studied in the main text.}
	\label{fig:4}
\end{figure}
Assuming for simplicity a standard polytropic equation of state between the pressure and the energy density,
\be
\rho = \frac{p}{K-1} + \eta_0 \lp \frac{p}{K \eta_0}\rp^{1/\Gamma}, \label{EOS}
\ee
where $\eta_0=16.6 \times 10^{12} {\rm g/cm}^3$,
one can simultaneously solve the Klein-Gordon and TOV set of equations to compute the solution of the scalar field in the interior of the star, from which one can extract the TLN using Eq.~\eqref{STLNNS}.
The result is shown in Fig.~\ref{fig:4} by the blue curve.
As expected, in the limit of small compactness the TLN tends to zero, as discussed above.
This can be understood by the fact that in the Newtonian limit ($C\to0$) the scalar perturbations do not feel any effective potential since the metric is flat.

\subsection{Axial vector TLNs}
The equation of motion for the axial decomposition of a spin-1 perturbation on the exterior geometry is given by
\begin{align}
	\lp 1- \frac{2 M}{r} \rp a'' +  \frac{2 M}{r^2} a' - 
	\frac{l(l+1)}{r^2} a  = 0.
\end{align}
For $l=2$, the exterior solution to this equation has the form
\begin{align}
& a_\text{\tiny ext} (r) = c_1 \frac{r^2 \lp 2 r - 3 M \rp}{16 M^3} + c_2 \llp \frac{2}{3 M^2} \lp 6 r^2 - 3 M r - M^2 \rp \right. \nonumber \\
& \left. + \frac{r^2(3 M - 2 r)}{ M^3} \log \lp \frac{r}{r-2 M} \rp \rrp,
\end{align}
whose behavior at asymptotic infinity $r \to \infty$ is given by
\be
a_\text{\tiny ext} (r) \to c_1\frac{1}{8} \frac{r^3}{M^3} - c_2\frac{4}{5} \frac{M^2}{r^2},
\ee
from which, using Eq.~\eqref{eq:Aphiexpansion}, one can extract the axial vector TLN
\begin{align}
	& k_2^{(s=1)} (y, C) = 3 C^5 (-1+2C) [6 + 3 C (y-2) -2y] \nonumber \\
	& \times \left\{
	10 C [-6(3+C(C-6)) +(2C-1)(-6 + C(C+3))y]\right. \nonumber \\
	& \left. + 15 (2C-1) [6 + 3 C (y-2) -2 y]\log(1-2C)
	\right\}^{-1}.
\end{align}
The equation of motion for the $l=2$ axial vector perturbation in the interior of the neutron star is given by~\cite{Cardoso:2017kgn}
\begin{align}
 e^{-\lambda} a'' - \lp \frac{2 r m' - r^2 \nu' +2 m (r \nu'-1)}{2 r^2} \rp a' \nonumber \ - \frac{6}{r^2} a  = 0.
\end{align}
In the Newtonian limit one has that
\be
k_2^{(s=1)} \to \frac{3-y}{4+2y} \qquad \text{when} \qquad C \to 0.
\ee
In this limit the interior solution of the vector perturbation goes as $a_\text{\tiny int} \sim c_1 r^3 + c_2 r^{-2}$, 
for which regularity at the origin implies $c_2 = 0$. This will result into $y \to 3$, for which $k \to 0$ as $C \to 0$, as in the scalar case.

In Fig.~\ref{fig:4} we show the $l=2$ axial vector TLN as a function of the stellar compactness by a red curve. As in the scalar case, the response goes to zero in the limit of small compactness because the effective potential for the vector perturbation vanishes in the Newtonian limit.
One can also appreciate that the vector TLN is larger than the scalar TLN, whereas it is smaller than the standard gravitational TLN~\cite{Hinderer:2007mb}, shown in Fig.~\ref{fig:4} by the green line. A striking difference in the tidal responses between the gravitational perturbation and the scalar or vector ones is given by the different asymptotic behaviour obtained in the limit $C \to 0$, which for the gravitational perturbation approaches the value $k_2^{(s=2)} \to 0.25991$~\cite{Hinderer:2007mb}.
This is due to the fact that $l=2$ gravitational perturbations involve also the quadrupolar perturbations of the fluid, which survive in the Newtonian limit, at variance with the scalar and vector TLNs that are purely due to spacetime curvature.

Finally, from Fig.~\ref{fig:4} one can deduce a hierarchy for the TLNs depending on the considered tidal perturbation, showing that the result for a gravitational perturbation is expected to be larger than the one obtained for scalar or vector perturbations.

\bibliography{draft}

\end{document}